\documentclass[a4paper,11pt,dvipsnames]{article}
\pdfoutput=1 %
\usepackage{jheppub} %
\usepackage{epsfig}
\usepackage{graphicx}
\usepackage{cancel}
\usepackage{color}
\usepackage{psfrag}
\usepackage{braket}
\usepackage{subfig}
\usepackage{amsfonts,amsmath,amssymb}
\usepackage[mathx]{mathabx}
\usepackage{slashed}
\usepackage[usenames,dvipsnames]{xcolor}
\usepackage{url}
\usepackage{longtable}
\usepackage{comment}

\newcommand{\nc}{\newcommand}

\newcommand{ \mysmall}[1]{\scriptscriptstyle #1} 

\newcommand{ \xt}{x_{\mysmall{t}}}

\newcommand{ \Op}{O}

\newcommand{ \cw}{\cos \theta_{\mysmall W}}
\newcommand{ \sw}{\sin \theta_{\mysmall W}}

\newcommand{ \sws}{\sin^2 \theta_{\mysmall W}}

\renewcommand{\Re}{{\rm Re \,}}

\renewcommand{\bar}[1]{\overline{#1}}
\renewcommand{\tilde}[1]{\widetilde{#1}}
\renewcommand{\check}[1]{\widecheck{#1}}

\newcommand{\f}{\frac}
\newcommand{\be }{\begin{equation}}   \nc{\ee }{\end{equation}}
\newcommand{\bea}{\begin{eqnarray}}   \nc{\eea}{\end{eqnarray}}
\newcommand{\baa}{\begin{array}}      \nc{\eaa}{\end{array}}
\newcommand{\bit}{\begin{itemize}}    \nc{\eit}{\end{itemize}}
\newcommand{\ben}{\begin{enumerate}}  \nc{\een}{\end{enumerate}}
\newcommand{\eps}{\varepsilon}

\def\ocal{{\cal O}}
\def\lcal{{\cal L}}

\newcommand{\eq}[1]{eq.~(\ref{#1})}
\newcommand{\Dlr}{\overset{\leftrightarrow}{D}}
\newcommand{\Dl}{\overset{\leftarrow}{D}}
\begin{document}
\preprint{\vbox{\hbox{CERN-PH-TH-2015-278}}}
\title{\boldmath Matching of gauge invariant dimension-six operators \\for $\mathbf{b\to s}$ and $\mathbf{b\to c}$ transitions  \bigskip}

\author[a]{Jason Aebischer} \emailAdd{aebischer@itp.unibe.ch}
\affiliation[a]{Albert Einstein Center for Fundamental Physics, Institute
  for Theoretical Physics,\\ University of Bern, CH-3012 Bern,
  Switzerland.}
\author[b]{Andreas Crivellin} \emailAdd{andreas.crivellin@cern.ch}
\affiliation[b]{CERN Theory Division, CH-1211 Geneva 23, Switzerland}
\author[a]{Matteo Fael}\emailAdd{fael@itp.unibe.ch}

\author[a]{Christoph Greub}\emailAdd{greub@itp.unibe.ch}

\abstract{New physics realized above the electroweak scale can be encoded in a model independent way in the Wilson coefficients of higher dimensional operators which are invariant under the Standard Model gauge group.
In this article, we study the matching of the $SU(3)_C \times SU(2)_L \times U(1)_Y$ gauge invariant dimension-six operators on the standard $B$ physics Hamiltonian relevant for $b \to s$ and $b\to c$ transitions. The matching is performed at the electroweak scale (after spontaneous symmetry breaking) by integrating out the top quark, $W$, $Z$ and the Higgs particle. We first carry out the matching of the dimension-six operators that give a contribution at tree level to the low energy Hamiltonian.
In a second step, we identify those gauge invariant operators that do not enter $b \to s$ transitions already at tree level, but can give relevant one-loop matching effects.}
\maketitle
\flushbottom

\section{Introduction \label{sec:intro}}

The Standard Model (SM) of particle physics as the gauge theory of strong and electroweak (EW) interactions has been tested and confirmed to a high precision since many years~\cite{PDG}.
Furthermore, the observation of a Higgs boson at the LHC~\cite{Aad:2012tfa,Chatrchyan:2012xdj} and the first measurements of its production and decay channels are consistent with the SM Higgs mechanism of EW symmetry breaking.

Nevertheless, the SM is expected to constitute only an effective theory valid up to a new physics (NP) scale $\Lambda$ where additional dynamic degrees of freedom enter. A renormalizable quantum field theory of NP, realized at a scale higher than the EW one, satisfies in general the following requirements:
\begin{itemize}
\item[({\it i})] Its gauge group must contain the SM gauge group~$SU(3)_C\times SU(2)_L\times U(1)_Y$ as a subgroup.
\item[({\it ii})] All SM degrees of freedom should be contained, either as fundamental or as composite fields.
\item[({\it iii})] At low-energies the SM should be reproduced, provided no undiscovered weakly coupled {\em light} particles exist (like axions or sterile neutrinos).
\end{itemize}

In most theories of physics beyond the SM that have been considered, the SM is recovered via the decoupling of heavy particles, with masses $\Lambda\gg M_Z$, guaranteed, at the perturbative level, by the Appelquist-Carazzone decoupling theorem~\cite{Appelquist:1974tg}. Therefore, NP can be encoded in higher-dimensional operators which are suppressed by powers of the NP scale $\Lambda$:
\be
 \label{eqn:Leff}
  \lcal_{\mathrm SM} =   \lcal_{\mathrm SM}^{(4)}   + \f{1}{\Lambda }  C_{\nu\nu}^{(5)} Q_{\nu\nu}^{(5)}
  + \f{1}{\Lambda^2} \sum_{k} C_k^{(6)} Q_k^{(6)}  + \ocal\left(\f{1}{\Lambda^3}\right)\,.
\ee
Here $\lcal_{\mathrm SM}^{(4)}$ is the usual renormalizable SM Lagrangian which contains only dimension-two and dimension-four operators, $Q_{\nu\nu}^{(5)}$ is the Weinberg operator giving rise to neutrino masses~\cite{Weinberg:1979sa}, $Q_k^{(6)}$ and $C_k^{(6)}$ denote the dimension-six operators and their corresponding Wilson coefficients, respectively~\cite{Buchmuller:1985jz,GIMR}.

Even if the ultimate theory of NP was not a quantum field theory, at low energies it would be described by an effective non-renormalizable Lagrangian~\cite{Weinberg:1995mt} and it would be possible to parametrize its effects at the EW scale in terms of the Wilson coefficients associated to these operators.
Thus, one can search for NP in a model independent way by studying the SM extended with higher-dimensional gauge-invariant operators. Once a specific NP model is chosen, the Wilson coefficients can be expressed in terms of the NP parameters by matching the beyond the SM theory under consideration on the SM enlarged with such higher dimensional operators.

Flavor observables, especially flavor changing neutral current processes, are excellent probes of physics beyond the SM: since in the SM they are suppressed by the Fermi constant $G_F$ as well as by small CKM elements and loop factors they are very sensitive to even small NP contributions. Therefore, on one hand flavor processes can stringently constrain the Wilson coefficients of the dimension-six operators induced by NP. On the other hand, if deviations from the SM were uncovered, flavor physics could be used as a guideline towards the construction of a theory of physics beyond the SM. The second point is especially interesting nowadays in light of the discrepancies between the SM predictions and the measurements of $b\to s\mu^+\mu^-$ and $b\to c\tau\nu$ processes: the combination of $B\to D^*\tau\nu$ and $B\to D\tau\nu$ branching fractions disagrees with the SM prediction~\cite{Fajfer:2012vx} at the level of 3.9 standard deviations ($\sigma$)~\cite{Amhis:2014hma}. Furthermore,  $b\to s\ell^+\ell^-$ global fits even show deviations between $4\,\sigma$ and $5\,\sigma$ \cite{Hurth:2014vma,Altmannshofer:2014rta,Descotes-Genon:2015uva}. These deviations have been extensively studied recently. Many NP models have been proposed to explain the anomalies, (see for example~\cite{Descotes-Genon:2013wba,Gauld:2013qba,Buras:2013qja,Gauld:2013qja,Buras:2013dea,Altmannshofer:2014cfa,Gripaios:2014tna,Crivellin:2015mga,Crivellin:2015lwa,Becirevic:2015asa,Niehoff:2015bfa,Varzielas:2015iva,Sierra:2015fma,Celis:2015ara,Sahoo:2015fla,Belanger:2015nma,Falkowski:2015zwa,Grinstein:2015aua,Gripaios:2015gra,Carmona:2015ena,Bauer:2015knc,Fajfer:2015ycq} for $b\to s\mu^+\mu^-$ and \cite{Crivellin:2012ye,Datta:2012qk,Celis:2012dk,Crivellin:2013wna,Li:2013vlx,Faisel:2013nra,Atoui:2013zza,Sakaki:2013bfa,Dorsner:2013tla,Biancofiore:2014wpa,Crivellin:2015hha,Freytsis:2015qca,Bauer:2015knc,Fajfer:2015ycq} for $b\to c\tau\nu$.). Therefore, at the moment, $B$ physics is probably our best guideline towards NP.

\begin{figure}[t]
  \centering
  \includegraphics[width=\textwidth]{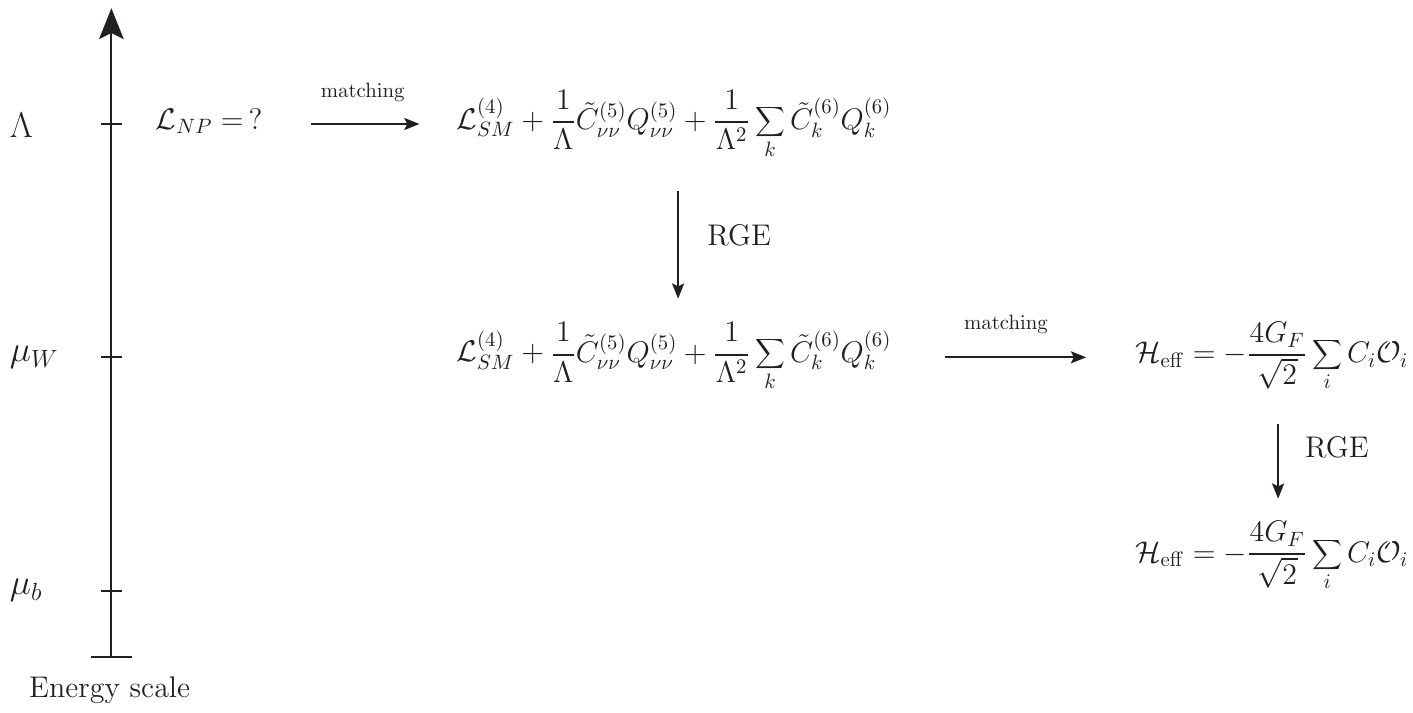}
  \caption{Mass scale hierarchy: the matching of the NP model onto the gauge invariant dimension-six operators is performed at the mass scale $\Lambda$. After the EW symmetry breaking, the matching of dimension-six operators on the effective Hamiltonian governing $B$~physics is performed at the mass scale $\mu_W$. The Wilson coefficients at different mass scales are connected via RGE evolution.
  }
  \label{fig:massscale}
\end{figure}

The effective field theory approach is an essential ingredient of all $B$ physics calculations within and beyond the SM. However, the Hamiltonian governing $b\to s$ and $b\to c$ transitions is not invariant under the full SM gauge group, but only under $SU(3)_C\times U(1)_{\rm EM}$ since it is defined below the EW scale where $SU(2)_L\times U(1)_Y$ is broken (see for example~\cite{Buchalla:1995vs,Buras:1998raa} for a review of the use of effective Hamiltonians in $B$ physics).
As a consequence, the SM extended with gauge invariant dimension-six operators must be matched onto the low energy effective Hamiltonian governing $B$ physics~(see figure~\ref{fig:massscale}).
In the flavor sector only partial analyses exist: the matching effects in the lepton sector were calculated in refs.~\cite{Crivellin:2013hpa,Crivellin:2014cta,Pruna:2014asa},\footnote{See ref.~\cite{Dassinger:2007ru} for an analysis of non-gauge invariant effective operators for tau decays.} while in the quark sector $b\to s\mu^+\mu^-$ transitions and their correlations  with $B\to K^{(*)}\nu\nu$ and $B\to D^{(*)}\tau\nu$ were studied in refs.~\cite{Bhattacharya:2014wla,Alonso:2015sja,Calibbi:2015kma,Alonso:2014csa,Buras:2014fpa}.
However a systematic and complete phenomenological study of the gauge invariant dimension-six operators in $B$ physics is still missing.
Such analysis proceeds, in a bottom-up approach, in the following three steps.
\begin{itemize}
  \item[({\it i})] The matching at the EW scale $\mu_W$, of the order of $M_W$, of the gauge invariant operators onto the low-energy $B$ physics Hamiltonian  by integrating out the heavy degrees of freedom represented by the top quark, the Higgs and the $Z$ and $W$ bosons.
    It is the aim of this article to perform such systematic matching of the gauge invariant operators.
  \item[({\it ii})] The evolution of the effective Hamiltonian's Wilson coefficients from the scale $\mu_W$ down to the $B$ meson scale $\mu_b$, where $\mu_b$ is of the order of $m_b$. This is obtained by solving the appropriate renormalizarion group equation (RGE)
    We note that after the matching procedure the set of operators in the $B$ physics Hamiltonian is larger than the SM one since new Lorentz structures must be taken into account, therefore the anomalous dimension matrices get also bigger compared to the SM.\footnote{For the anomalous dimension matrices beyond the SM for $\Delta F=2$ processes see for example refs.~\cite{Ciuchini:1997bw,Buras:2000if}, for 4-fermion operators ref.~\cite{Buras:2000if} and for $b\to s\gamma$ refs.~\cite{Borzumati:1998tg,Borzumati:1999qt}.}
  \item[({\it iii})] The assessment of the constrains on the dimension-six operators' Wilson coefficients (defined at the EW scale $\mu_W$) stemming from the available flavor observables. An example of such analysis can be found in the section~\ref{sec:example}, while the complete numerical analysis will be given in a subsequent publication.
\end{itemize}
The purpose of the outlined study is to depict the general pattern of deviations observed in $B$ physics employing dimension-six operators.
It is worth noting however that in the framework of higher dimensional operators, in order to correctly interpret  any deviations of the SM  in terms of a specific NP model, it is necessary to map the pattern of deviations observed at the EW scale back to the scale $\Lambda$ where the BSM physics was supposedly integrated out (see figure\ref{fig:massscale}).
Indeed due to operator mixing, the pattern of deviations at the EW scale differs from the pattern of Wilson coefficients at the matching scale $\Lambda$. The connection between these two mass scales is given by the RGE evolution of dimension-six operators~\cite{Jenkins:2013zja,Jenkins:2013wua,Alonso:2013hga}.

The outline of this article is as follow: in section~\ref{sec:operators} we list the operators relevant for our analysis and discuss the EW symmetry breaking. Then, in section~\ref{sec:tree}, we establish our conventions for the $B$ physics Hamiltonian and we perform the complete matching of the dimension-six operators that give contributions to $b\to s$ or $b\to c$ transitions at tree level. In section~\ref{sec:1-loop} we identify and calculate the leading one-loop EW matching corrections for $b \to s$ processes for those operators which do not enter $b \to s$ transitions already at tree-level. A phenomenological application of the computed matching conditions will be given in section~\ref{sec:example}. Finally we conclude.

\section{Gauge invariant operators}
\label{sec:operators}

In this section we list the gauge invariant operators, following the conventions of ref.~\cite{GIMR}, that contribute to $b\to s$ or $b\to c$ transitions at tree-level. Here we only consider operators involving quark fields. The importance of flavor physics in constraining operators which modify triple gauge couplings has been studied in ref.~\cite{Bobeth:2015zqa}. Recall that the gauge invariant dimension-six operators are defined before EW symmetry breaking, implying that they are given in the interaction basis (as the mass basis it not yet defined). After the EW symmetry breaking, the fermions acquire their masses and the necessary diagonalizations of their mass matrices affect the Wilson coefficients. As we will see, all these rotations can be absorbed by a redefinition of the Wilson coefficients, except for the misalignment between the left-handed up-quark and down-quark rotations, i.e.\ the Cabibbo-Kobayashi-Maskawa matrix (CKM) which relates charged and neutral currents.

\subsection{Operator formalism}
In table~\ref{tab:operators1} we list the operators contributing to $b\to s$ at the tree level (and possibly also to $b\to c$ transitions), while table~\ref{tab:operators3} gives the operators generating at tree level $b \to c$ but not $b\to s$.
For the SM Lagrangian we adopt the standard definition
\begin{multline}
  \mathcal{L}_{SM}^{(4)} =
  -\frac{1}{4} G_{\mu\nu}^{ A} G^{{ A} \mu\nu}
  -\frac{1}{4} W_{\mu\nu}^{ I} W^{{I} \mu\nu}
  -\frac{1}{4} B_{\mu\nu} B^{\mu\nu}
  +(D_\mu \varphi)^\dagger (D^\mu \varphi)
  +m^2 \varphi^\dagger \varphi
  -\frac{1}{2} \lambda \left( \varphi^\dagger \varphi \right)^2 \\
  +i \Big( \bar{\ell} \slashed D \ell + \bar{e}\slashed D e
  +\bar{q}\slashed D q +  \bar{u}\slashed D u + \bar{d}\slashed D d  \Big)
  -\Big(  \bar{\ell} \, Y_e e \, \varphi
  + \bar{q} \, Y_u u \, \tilde{\varphi}
  + \bar{q} \, Y_d d \, \varphi + {\rm h.c.} \Big) \, ,
  \label{eqn:SMLagrangian}
\end{multline}
where $\ell$, $q$ and $\varphi$ stand for the lepton, quark and Higgs $SU(2)_L$ doublets, respectively, while the right-handed isospin singlets are denoted by $e$, $u$ and $d$.
Here $\tilde{\varphi}^i = \varepsilon_{ij} (\varphi^j)^*$, where $\varepsilon_{ij}$ is the totally antisymmetric tensor with $\varepsilon_{12} = +1$.
Flavor indices $i,j,k,l =1,2,3$ are implicitly assigned to each fermion field appearing in~\eqref{eqn:SMLagrangian}, and the Yukawa couplings $Y_{e,u,d}$ are matrices in the generation space. Therefore, in table~\ref{tab:operators1} the operator names in the left column of each block should be supplemented with generation indices of the fermion fields whenever necessary.
Covariant derivatives are defined with the plus sign, i.e.\ for example
\begin{equation}
  D_\mu q = (\partial_\mu
  + i g_s T^A G^A_\mu
  + i g \frac{\tau^I}{2} W^I_\mu
  +i g' Y B_\mu) \, q,
\end{equation}
where $Y$ is the hypercharge and $T^A=\frac{1}{2}\lambda^A$; $\lambda^A$ and $\tau^I$ are the Gell-Mann and Pauli matrices, respectively. With the above definition for the covariant derivative, the gauge field strength tensors read
\begin{align}
  G_{\mu\nu}^A &= \partial_\mu G_\nu^A-\partial_\nu G_\mu^A-g_s f^{ABC} G_\mu^B G_\nu^C, \notag \\
  W_{\mu\nu}^I &= \partial_\mu W_\nu^I-\partial_\nu W_\mu^I-g \,\varepsilon^{IJK} W_\mu^J W_\nu^K, \notag \\
  B_{\mu\nu} &= \partial_\mu B_\nu-\partial_\nu B_\mu.
\end{align}
Moreover the Hermitian derivative terms are defined as
\begin{align}
  \varphi^\dagger \, i \Dlr_\mu  \varphi &=
  i \varphi^\dagger (D_\mu - \Dl_\mu) \varphi,&
  \varphi^\dagger \, i \Dlr_\mu \! ^{I} \varphi &=
  i \varphi^\dagger (\tau^I D_\mu - \Dl_\mu \tau^I) \varphi,&
\end{align}
where $\varphi^\dagger \Dl_\mu \varphi = (D_\mu \varphi)^\dagger \varphi$.
For further details concerning conventions and notations, we refer the reader to ref.~\cite{GIMR}.

For the operators in the classes $(\bar{L}L)(\bar{L}L), (\bar{L}L)(\bar{R}R), (\bar{R}R)(\bar{R}R)$ and $\psi^2 \varphi^2 D$ (except for $Q_{\varphi ud}$), hermitian conjugation is equivalent to the transposition of generation indices in each of the fermion currents. Moreover, the operators $Q_{qq}^{(1)}, Q_{qq}^{(3)},Q_{uu}$ and $Q_{dd}$ are symmetric under exchange of the flavor indices $ij \leftrightarrow kl$. Therefore, we will restrict ourselves to the operators satisfying $[ij]<[kl]$, where $[ij]$ denotes the two digit number $[ij] = 10 i +j$.

\begin{table}[t]
  \centering
  \renewcommand{\arraystretch}{1.2}
  \small
  \begin{tabular}{||c|c|c|c|c|c||}
    \hline \hline
    \multicolumn{2}{||c|}{$(\bar{L}R)(\bar{R}L)$ or $(\bar{L}R)(\bar{L}R)$}&
    \multicolumn{2}{|c|}{$(\bar{L}L)(\bar{L}L)$} &
    \multicolumn{2}{|c||}{$\psi^2 X \varphi$} \\
    \hline
    $Q_{\ell edq}$ &
    $(\bar{\ell }_i^a e_j) (\bar{d}_k q_l^a)$ &
    $Q_{qq}^{(1)}$  &
    $(\bar q_i \gamma_\mu q_j)(\bar q_k \gamma^\mu q_l)$ &
    $Q_{dW}$ &
    $( \bar{q}_i \sigma^{\mu\nu} d_j ) \tau^{\mysmall I} \varphi W^{\mysmall I}_{\mu\nu}$ \\
    $Q_{quqd}^{(1)}$ &
    $(\bar q^a_i u_j) \eps_{ab} (\bar q^b_k d_l)$ &
    $Q_{\ell q}^{(1)}$                &
    $(\bar \ell_i \gamma_\mu \ell_j)(\bar q_k \gamma^\mu q_l)$ &
    $Q_{dB}$ &
    $( \bar{q}_i \sigma^{\mu\nu} d_j ) \varphi B_{\mu\nu}$ \\
    $Q_{quqd}^{(8)}$ &
    $(\bar q^a_i T^{\mysmall A}u_j) \eps_{ab} (\bar q^b_k T^{\mysmall A}d_l)$   &
    $Q_{qq}^{(3)}$  &
    $(\bar q_i \gamma_\mu \tau^I q_j)(\bar q_k \gamma^\mu \tau^I q_l)$ &
    $Q_{dG}$ &
    $(\bar{q}_i \sigma^{\mu\nu} T^{\mysmall A}d_j) \varphi G^{\mysmall A}_{\mu\nu}$ \\\cline{1-2}\cline{5-6}
    \multicolumn{2}{||c|}{$(\bar{L}L)(\bar{R}R)$} &
    $Q_{\ell q}^{(3)}$                &
    $(\bar \ell_i \gamma_\mu \tau^I \ell_j)(\bar q_k \gamma^\mu \tau^I q_l)$  &
   \multicolumn{2}{|c||}{$\psi^2 \varphi^3 $} \\\cline{1-2}\cline{5-6}
    $Q_{\ell d}$ &
    $(\bar \ell_i \gamma_\mu \ell_j)(\bar d_k \gamma^\mu d_l)$ &
    & &
    $Q_{d\varphi}$  &
    $(\varphi^{\dag}\varphi)(\bar q_i\,d_j\,\varphi)$  \\ \cline{3-6}
    $Q_{qe}$    &
    $(\bar q_i \gamma_\mu q_j)(\bar e_k \gamma^\mu e_l)$ &
    \multicolumn{2}{|c|}{$(\bar{R}R)(\bar{R}R)$} &
    \multicolumn{2}{|c||}{$\psi^2 \varphi^2 D $}
    \\ \cline{3-6}
    $Q_{qu}^{(1)}$         &
    $(\bar q_i \gamma_\mu q_j)(\bar u_k \gamma^\mu u_l)$ &
    $Q_{dd}$        &
    $(\bar d_i \gamma_\mu d_j)(\bar d_k \gamma^\mu d_l)$ &
    $Q_{\varphi q}^{(1)}$ &
    $( \varphi^\dagger i \Dlr_\mu \varphi )
    ( \bar{q}_i \gamma^\mu q_j)$ \\
    $Q_{qd}^{(1)}$         &
    $(\bar q_i \gamma_\mu q_j)(\bar d_k \gamma^\mu d_l)$ &
    $Q_{ed}$  &
    $(\bar e_i \gamma_\mu e_j)(\bar d_k \gamma^\mu d_l)$ &
    $Q_{\varphi q}^{(3)}$ &
    $( \varphi^\dagger i \Dlr_\mu \! ^{\mysmall I} \varphi )
    ( \bar{q}_i \tau^{\mysmall I} \gamma^\mu q_j )$ \\
    $Q_{qu}^{(8)}$         &
    $(\bar q_i \gamma_\mu T^{\mysmall A} q_j)(\bar u_k \gamma^\mu T^{\mysmall A} u_l)$ &
    $Q_{ud}^{(1)}$                &
    $(\bar u_i \gamma_\mu u_j)(\bar d_k \gamma^\mu d_l)$ &
    $Q_{\varphi d}$ &
    $( \varphi^\dagger i \Dlr_\mu \varphi )
    ( \bar{d}_i \gamma^\mu d_j )$    \\
    $Q_{qd}^{(8)}$         &
    $(\bar q_i \gamma_\mu T^{\mysmall A} q_j)(\bar d_k \gamma^\mu T^{\mysmall A} d_l)$ &
    $Q_{ud}^{(8)}$                &
    $(\bar u_i \gamma_\mu T^{\mysmall A}u_j)(\bar d_k \gamma^\mu T^{\mysmall A} d_l)$ &
    $Q_{\varphi ud}$ &
    $ i( \tilde{\varphi}^\dagger D_\mu \varphi)(\bar{u}_i \gamma^\mu d_j) $\\
  \hline \hline
  \end{tabular}
  \caption{Complete list of the dimension-six operators that contribute to
  $b \to s$ (and possibly also to $b \to c$) transitions at tree level.}
  \label{tab:operators1}
\end{table}
\begin{table}[t]
  \centering
  \renewcommand{\arraystretch}{1.2}
  \begin{tabular}{||c|c||}
    \hline \hline
    \multicolumn{2}{||c||}{$(\bar{L}R)(\bar{L}R)$}\\
    \hline
    $Q_{\ell equ}^{(1)}$ &
    $(\bar{\ell}_i^ae_j)\varepsilon_{ab}(\bar{q}^b_k u_l)$\\
    $Q_{\ell equ}^{(3)}$ &
    $(\bar{\ell}_i^a \sigma^{\mu\nu} e_j)\varepsilon_{ab}(\bar{q}^b_k \sigma_{\mu\nu} u_l)$\\
    \hline \hline
  \end{tabular}
  \caption{The two dimension-six operators that contribute to $b \to c$ but not to $b\to s$ transitions at tree level.}
  \label{tab:operators3}
\end{table}

\subsection{EW symmetry breaking}

Although the set of gauge invariant dimension-six operators we have just introduced is written in term of the flavor basis, actual calculations that confront theory with experiment are performed using the mass  eigenbasis which is defined after the EW symmetry breaking. In the broken phase, flavor and mass eigenstates are not identical and the $SU(2)_L$ doublet components are distinguishable. Therefore, we need to rotate the weak eigenstates into mass eigenstates via the following transformations:
\begin{align}
  u_L^{i} &\to S^{u}_{L\; ij} \; u_L^j,& u_R^{i} &\to S^{u}_{R \; ij} \, u_R^j\,,\\
  d_L^{i} &\to S^{d}_{L\; ij} \; d_L^j,& d_R^{i} &\to S^{d}_{R \; ij} \, d_R^j\,,
  \label{eqn:flavormassbasis}
\end{align}
where $S^d_L,S^d_R,S^u_L$ and $S^u_R$ are the $3 \times 3$ unitary matrices in flavor space that diagonalize the mass matrix as
\begin{equation}
	S^{q\dagger}_{L \, ii'}\, m_q^{i'j'} S^q_{R\, j'j} = m_{q_i} \delta_{ij}\,.
\end{equation}
With these definitions, the CKM matrix $V$ is given by
\begin{equation}
  V = \left( S_L^u \right)^\dagger S_L^d\,.
  \label{eqn:defCKm}
\end{equation}

After these necessary field redefinitions, there are no flavor changing neutral currents at tree-level in the SM, due to the unitarity of the transformation matrices, and mixing between generations only occurs in the charged quark current. When dimension-six operators are included in the Lagrangian, the effect on them by the matrices $S_{L,R}^q$ cannot be eliminated by unitarity. However, these rotations can be absorbed into the Wilson coefficients. As a first example, we consider the operator $Q_{\varphi d}$ which takes the form:
\begin{equation}
  C_{\varphi d}^{mn} Q_{\varphi d}^{mn} =
  C_{\varphi d}^{mn}
  \left( \varphi^\dagger i \overset{\leftrightarrow}{D}_\mu \varphi \right)
  \left( \bar{d}_R^{m} \gamma^\mu d_R^{n} \right)
  \to   C_{\varphi d}^{mn}
  \left( \varphi^\dagger i \overset{\leftrightarrow}{D}_\mu \varphi \right)
  \left( \bar{d}_R^{i} S^{d\dagger}_{R \, im} \gamma^\mu S^d_{R \, nj} d_R^{j} \right)\,.
\end{equation}
Redefining
\begin{equation}
	\tilde{C}_{\varphi d}^{ij} = C_{\varphi d}^{mn} S^{d\dagger}_{R \, im} S^d_{R \, nj} \,,
\end{equation}
we can indeed absorb $S_{L,R}^q$ into the overall coefficient:
\begin{equation}
  C_{\varphi d}^{mn} Q_{\varphi d}^{mn} =
  \tilde{C}_{\varphi d}^{ij}
  \left( \varphi^\dagger i \overset{\leftrightarrow}{D}_\mu \varphi \right)
  \left( \bar{d}_R^{i} \gamma^\mu d_R^{j} \right)\,.
\end{equation}
In contrast to the SM, it is not possible anymore to avoid the appearance of flavor changing neutral currents for all operators. Moreover, the redefinitions of the Wilson coefficients are not unique, in general. Let us consider as a second example the operator $Q_{\varphi q}^{(1)}$:
\begin{align}
  C_{\varphi q}^{(1)\,mn} Q_{\varphi q}^{(1)\,mn} &=
  C_{\varphi q}^{(1)\,mn}
  \left( \varphi^\dagger i \overset{\leftrightarrow}{D}_\mu \varphi \right)
  \left( \bar{u}^{m}_L \gamma^\mu u^{n}_L + \bar{d}^{m}_L \gamma^\mu d^{n}_L \right) \\
  &\to   C_{\varphi q}^{(1)\,mn}
  \left( \varphi^\dagger i \overset{\leftrightarrow}{D}_\mu \varphi \right)
  \left( \bar{u}^{i}_L S^{u\dagger}_{L \, im} \gamma^\mu S^u_{L \, nj} u^{j}_L
  +\bar{d}^{i}_L S^{d\dagger}_{L \, im} \gamma^\mu S^d_{L \, nj} d^{j}_L\right)\,.
\end{align}
In this case we cannot absorb at the same time the rotation for the up quarks and for the down quarks, so that we can choose to define
\begin{equation}
	\tilde{C}_{\varphi q}^{(1)\,ij} =
C_{\varphi q}^{(1)\,mn} S^{d\dagger}_{L \, im} S^d_{L \, nj}\,,
\end{equation}
or
\begin{equation}
	\check{C}_{\varphi q}^{(1)\,ij} =
C_{\varphi q}^{(1)\,mn} S^{u\dagger}_{L \, im} S^u_{L \, nj} \,,
\end{equation}
obtaining the two equivalent expressions
\begin{align}
   C_{\varphi q}^{(1)\,mn} Q_{\varphi q}^{(1)\,mn} &=
   \tilde{C}_{\varphi q}^{(1)\,ij}
  \left( \varphi^\dagger i \overset{\leftrightarrow}{D}_\mu \varphi \right)
  \left( V_{ki} V^*_{lj} \bar{u}^{k}_L \gamma^\mu u^{l}_L
  +\bar{d}^{i}_L \gamma^\mu d^{j}_L\right)\\
  &= \check{C}_{\varphi q}^{(1)\,ij}
  \left( \varphi^\dagger i \overset{\leftrightarrow}{D}_\mu \varphi\right)
  \left( \bar{u}^{i}_L \gamma^\mu u^{j}_L
  + V^*_{ik} V_{jl} \bar{d}^{k}_L \gamma^\mu d^{l}_L\right)\,.
\end{align}
For both definitions, the mass diagonalization leads to flavor changing neutral currents either in the up sector, for the coefficient denoted with the tilde ($\sim$), or in the down sector for that one with the check ($\vee$). The two notations are related through the identity
\begin{equation}
	\check{C}^{ij} = V_{ik} V^*_{jl} \tilde{C}^{kl}\,.
\end{equation}
All operators reported in table~\ref{tab:operators1} must be analogously expressed in the mass basis. We report in appendix~\ref{sec:appop} the explicit expressions for the Wilson coefficients $\tilde{C}$.

\subsection[$Q_{d\varphi}$ and $Q_{u\varphi}$]
{\boldmath $Q_{d\varphi}$ and $Q_{u\varphi}$}

The operators $Q_{d\varphi}$ and $Q_{u\varphi}$ play a special role as they contribute to the quark mass matrices after the EW symmetry breaking. For example, the down-quark mass matrix receives two contributions, one from the SM Yukawa interactions and one from the operator $Q_{d\varphi}$:
\begin{equation}
    m^{ij}_d =
    \frac{v}{{\sqrt 2 }}
    \left( {Y^{ij}_d - \frac{1}{2}\frac{{{v^2}}}{{{\Lambda ^2}}}C_{d\varphi}^{ij}}
    \right)\,, \label{Hdd-mass}
\end{equation}
where $Y_d$ is the Yukawa matrix of the SM and $v=246$ GeV is the vacuum expectation value of the SM Higgs field. For the coupling of the Higgs with the down-type quarks, defined by the Lagrangian term $\mathcal{L}_H= - h \, \bar{d}_L \Gamma^h d_R+ \mathrm{ h.c.}$, the extra contribution is enhanced by a combinatorial factor of three compared to the contribution to the mass term:
\begin{equation}
 \Gamma^h_{d_id_j}=
 \frac{1}{\sqrt{2}}
 \left(Y_d^{ij} -\frac{3}{2} \frac{v^2}{\Lambda^2} C^{ij}_{d\varphi} \right)
 = {\frac{{m^{ij}_d }}{v}
 - \frac{1}{{\sqrt 2 }}\frac{{{v^2}}}{{{\Lambda ^2}}}C_{d\varphi}^{ij}} \,.
 \label{eqn:Hdd-coupling}
\end{equation}
Unlike in the pure dimension-four SM, the mass matrix and the quark-Higgs coupling cannot be diagonalized simultaneously: a flavor changing interaction between the SM Higgs and the quarks appears~\cite{Blankenburg:2012ex,Davidson:2012ds,Crivellin:2014cta}. Indeed the first term in~\eq{eqn:Hdd-coupling} is rendered diagonal by a field redefinition as in~\eqref{eqn:flavormassbasis},
\begin{equation}
U_{L \, ii'}^{d\dag }\, m^{i'j'}_d U_{R \, j'j}^{d} = m_{{d_i}}^{}{\delta _{ij}}\,,
\end{equation}
where the new $U_{L,R}^d$ matrices, necessary to diagonalize the mass in the presence of the $Q_{d\varphi}$ operator, differ from $S_{L,R}^d$ by terms of order $1/\Lambda^2$. The quark-Higgs coupling matrix is now given by
\begin{equation}
\Gamma^h_{d_id_j}=\frac{m_{d_i}}{v} \delta_{ij} -\frac{1}{\sqrt{2}} \frac{v^2}{\Lambda ^2}  \tilde{C}_{d\varphi}^{ij}\,,
\end{equation}
where we have defined
\begin{equation}
	\tilde C_{d\varphi}^{ij} =	\left( {U_L^{d\dag }C_{d\varphi}U_R^d} \right)_{ij}
	= \left(S^{d \dagger}_L C_{d\varphi} S^d_R\right)_{ij} 	+ O\left(\frac{1}{\Lambda^2} \right) \,.
\end{equation}
Note that in this approximation all Wilson coefficients of the operators discussed above remain unchanged since the extra rotation induced by the $Q_{d\varphi}$ operator would lead to a $1/\Lambda^4$ effect. Similar considerations apply to the operator $Q_{u\varphi}$.

\section{Tree level matching}
\label{sec:tree}

In this section we perform the tree-level matching of the gauge invariant dimension-six operators relevant for $b\to s$ and $b\to c$ transitions. This matching is performed at the EW scale on the effective Hamiltonian governing $B$ physics, which is defined below the EW scale. Therefore, the effective $B$ physics Hamiltonian contains the SM fields without $W$, $Z$, the Higgs and the top quark, while these are dynamical fields of the gauge invariant dimension-six operator basis. As we will see, the $B$-physics Hamiltonian contains operators with additional Lorentz structures compared to the ones relevant in the SM.

\subsection[$\Delta B=\Delta S=2$]{\boldmath $\Delta B=\Delta S=2$}

In this section we consider $B_s$-$\overline{B}_s$ mixing. Here, following the conventions of refs.~\cite{Ciuchini:1997bw,Becirevic:2001jj}, the effective Hamiltonian is given by

\begin{equation}
{\cal H}^{\mysmall \Delta {B}=\Delta S=2}_{\mysmall \rm eff}=\sum\limits_{j = 1}^5 C_j \, O_j  +
\sum\limits_{j = 1}^3 C^\prime_j \, O^\prime_j + {\rm h.c.} \, ,
\label{eqn:Heff_DeltaF2}
\end{equation}
with the operators defined as
\begin{align}
O_1   \,= & \, ({\bar s} \gamma_{\mu} P_{L} b)\, ({\bar s} \gamma^{\mu} P_{L} b)\,, &
O_2  \,= & \, ({\bar s}P_{L} b)\, ({\bar s}  P_{L} b)     \,,   \notag \\
O_3  \,= &\, ({\bar s}_{\alpha} P_{L} b_{\beta})\, ({\bar s}_{\beta}  P_{L} b_{\alpha})     \,,                &
O_4   \,= & \, ({\bar s} P_{L} b)\, ({\bar s}  P_{R} b)     \,,  \notag  \\
O_5  \,= & \,({\bar s}_{\alpha} P_{L} b_{\beta})\, ({\bar s}_{\beta}  P_{R} b_{\alpha})     \, ,                &
\label{opbasisDF2}
\end{align}
where $\alpha$ and $\beta$ are color indices. The primed operators $O_{1,2,3}^\prime$ are obtained from $O_{1,2,3}$ by interchanging $P_L$ with $P_R$.

The contributions from the four-fermion operators to the Hamiltonian in eq.~\eqref{eqn:Heff_DeltaF2} read:
\begin{align}
  C_1 &= -\frac{1}{\Lambda^2}  \left[ \tilde{C}_{qq}^{(1) \, 2323} + \tilde{C}_{qq}^{(3) \, 2323} \right]\,,\\
  C_1' &= -\frac{1}{\Lambda^2}   \tilde{C}_{dd}^{2323} \,,\\
  C_4 &=  \frac{1}{\Lambda^2} \tilde{C}_{qd}^{(8)\, 2323}\,,\\
  C_5 &=  \frac{1}{\Lambda^2} \left[2 \tilde{C}_{qd}^{(1)\, 2323}- \frac{1}{N_c}\tilde{C}_{qd}^{(8)\, 2323}  \right]\,,
\end{align}
where $N_c$ denotes the number of colors.
In addition, we include for completeness the effects of $Q_{d\varphi}$ even though they are formally suppressed by $1/\Lambda^4$ because the $1/\Lambda^2$ effect in the $B$-physics Hamiltonian is suppressed due to the $m_f/v$ coupling of the Higgs to the light fermions.\footnote{Note that this counting argument already suggest, that the EFT approach to flavor changing Higgs decays has quite limited applicability.}
Here we get
\begin{align}
C_2^{} &=-\frac{{ 1}}{{2m_h^2}}{\mkern 1mu} {\left( {\Gamma _{bs}^{{\kern 1pt} h * }} \right)^2}\,,\\
C_2^{\prime} &=-\frac{{  1}}{{2m_h^2}}{\mkern 1mu} {\left( {\Gamma _{sb}^h} \right)^2}\,,\\
C_4^{\kern 1pt}  &=-\frac{{ 1}}{{m_h^2}}{\mkern 1mu} \Gamma _{sb}^h{\mkern 1mu} \Gamma _{bs}^{h * }\,,
\end{align}
where $\Gamma^h_{d_i d_j}$ is defined in eq.~\eqref{eqn:Hdd-coupling}. Note that we do not include the analogous contributions from a modified $Z$ coupling since in this case the coupling to light fermions are not suppressed and especially $b\to s\mu^+\mu^-$ processes will give relevant tree-level constraints at the $1/\Lambda^2$ level.

\subsection[$\Delta B=\Delta C=1$]{\boldmath $\Delta B=\Delta C=1$}

For the charged current process $b\to c \ell_i\nu_j$ we write the effective Hamiltonian as
\begin{equation}
  \mathcal{H}_{\mysmall \rm eff}^{\mysmall \Delta B= \Delta C =1} =
  -\frac{4G_F}{\sqrt{2}}  \left[
   C_T \, \Op_T+
  \sum_{\scriptstyle i=S,V}  C_i \, \Op_i+ C_i' \,\Op_i'
  \right],
  \label{eqn:Heff_DeltaB1DeltaC1}
\end{equation}
where the operators are
\begin{align}
\Op_V &= \left( \bar{c} \, \gamma^\mu P_L b \right)  \left( \bar{\ell} \, \gamma_\mu P_L \nu \right)\,, &
\Op_T &= \left( \bar{c} \, \sigma^{\mu\nu} P_L b \right)  \left( \bar{\ell} \, \sigma_{\mu\nu} P_L \nu \right)\,, &
\Op_S &= \left( \bar{c}\, P_L b \right)  \left( \bar{\ell} \, P_L \nu \right)\,,
\end{align}
and the prime operators are obtained by interchanging $P_L \leftrightarrow P_R$ in the quark current.\footnote{The operator $\Op_T'$ is identically zero due to Fierz transformations.}

The four-fermion operators lead to the following contribution to the effective Hamiltonian:
\begin{align}
  C_V &= \frac{v^2}{\Lambda^2} V_{ci}\, \tilde{C}^{(3) \, ll i3}_{\ell q}  \,,&
  C_S' &= \frac{v^2}{2\Lambda^2} V_{ci} \, \tilde{C}^{* \, ll3i}_{\ell edq}\,,\\
  C_S &= \frac{v^2}{2\Lambda^2} V_{ib}\,\tilde{C}^{* (1)\, lli2}_{\ell equ}\,,&
  C_T &= \frac{v^2}{2\Lambda^2} V_{ib}\,\tilde{C}^{* (3)\, lli2}_{\ell equ}\,,
\end{align}
where the summation over $i=1,2,3$ is understood. The operators  $Q_{\varphi ud}$ and $Q_{\varphi q}^{(3)}$ induce an anomalous $u$-$d$-$W$ coupling. Their contribution to the $b \to c \ell \nu$ transition reads:

\begin{align}
   C_V^\prime &=    -\frac{v^2}{2\Lambda^2}   \tilde{C}_{\varphi ud}^{23}\,, \\
   C_V &=  -\frac{v^2}{\Lambda^2}   V_{ci}  \tilde{C}_{\varphi q}^{(3) i3} \,.
 \end{align}
The effect of such modified $W$ couplings to quarks on the determination of $V_{cb}$ (and analogously on $V_{ub}$) has been discussed in refs.~\cite{Voloshin:1997zi,Dassinger:2007pj,Chen:2008se,Dassinger:2008as,Crivellin:2009sd,Cirigliano:2009wk,He:2009hz,Buras:2010pz,Feger:2010qc,Crivellin:2014zpa,Bernlochner:2014ova}.

In principle, also momentum dependent modifications of the $W$-$c$-$b$ coupling can lead to effects in $b\to c \ell\nu$ transitions as examined in refs.~\cite{Dassinger:2008as,Feger:2010qc} at the level of non-gauge invariant operators. However, these effects scale like $m_bv/(m_W^2 \Lambda^2)$. Furthermore, also corrections to $Z$-$b$-$b$ couplings can appear which are stringently constrained, making the possible contributions tiny~\cite{Crivellin:2014zpa}. Therefore we do not include these effects here.


\subsection[$\Delta B=\Delta S=1$]
{\boldmath $\Delta B=\Delta S=1$}
We describe the $b \to s \ell^- \ell^{\prime+}$ and $b \to s \gamma$ transition via the effective Hamiltonian
\begin{equation}
  \mathcal{H}_{\mysmall \rm eff}^{\mysmall \Delta B = \Delta S =1} =
  -\frac{4G_F}{\sqrt{2}}
  \left(   \sum_{i}   C_i\, \Op_i+C'_i\, \Op'_i
  +\sum_{i} \sum_q  C_i^q \, \Op_i^q +C_i^{'q} \, \Op_i^{'q} \right)\,,
  \label{eqn:Heff_DeltaB1DeltaS1}
\end{equation}
where the index $q$ runs over all light quarks $q=u,d,c,s,b$. The operators contributing in the first part are:
\begin{align}
  \Op_1 &= (\bar{s} \, T^{\mysmall A} \gamma_\mu P_L c) \;
  (\bar{c} \, T^{\mysmall A} \gamma^\mu P_L b)\,,&
  \Op_2 &=(\bar{s} \gamma_\mu P_L  c) \;
  (\bar{c} \gamma^\mu P_L b),\notag\\
  \Op_7 &= \frac{e}{16\pi^2} m_b \,  ( \bar{s}  \, \sigma_{\mu\nu} P_R \,  b ) \; F^{\mu\nu},&
  \Op_8 &= \frac{g_s}{16\pi^2} m_b \,
  ( \bar{s}\, T^{\mysmall A} \sigma_{\mu\nu} P_R \, \, b ) \; G^{\mu\nu \, \mysmall A}\,,\notag \\
  \Op^{\ell\ell^\prime}_9 &= \frac{e^2}{16 \pi^2}  (\bar{s}\,\gamma_\mu P_L b) \;
    (\bar{\ell} \gamma^\mu \ell^\prime)\,, &
  \Op^{\ell\ell^\prime}_{10} &=\frac{e^2}{16 \pi^2}
   (\bar{s} \gamma_\mu P_L b) \;    (\bar{\ell} \gamma^\mu \gamma_5 \ell^\prime)\,, \label{eqn:O10}\notag \\
  \Op^{\ell\ell^\prime}_S &= (\bar{s} P_R b) \; (\bar{\ell} \ell^\prime)\,,&
  \Op_P^{\ell \ell'} &= (\bar{s} P_R b) \; (\bar{\ell} \gamma_5 \ell^\prime)\,,\notag\\
  \Op^{\ell\ell^\prime}_T & =(\bar{s} \sigma^{\mu\nu} b) \; (\bar{\ell} \sigma_{\mu\nu} \ell^\prime)\,,&
  \Op^{\ell\ell^\prime}_{T5} & =(\bar{s} \sigma^{\mu\nu} b) \; (\bar{\ell} \sigma_{\mu\nu} \gamma_5 \ell^\prime)\,.
\end{align}
While in the second part of the Hamiltonian we have four-quark operators with vectorial Lorentz structures,
\begin{align}
  \Op_3^q &= (\bar{s} \gamma_\mu P_Lb)\;   ( \bar{q} \gamma^\mu q ) &
  \Op_4^q &=(\bar{s}\, T^{\mysmall A} \gamma_\mu P_L b)\;
  (\bar{q}\, T^{\mysmall A} \gamma^\mu q )\,, \notag\\
  \Op_5^q&=(\bar{s}\gamma_\mu \gamma_\nu \gamma_\rho P_L b)\;
  ( \bar{q} \gamma^\mu \gamma^\nu \gamma^\rho q)\,, &
  \Op_6^q &=(\bar{s}\, T^{\mysmall A}\gamma_\mu \gamma_\nu \gamma_\rho P_L b)\;
  (\bar{q}\, T^{\mysmall A} \gamma^\mu \gamma^\nu \gamma^\rho q)\,,
\end{align}
and four-quark operators with scalar and tensor Lorentz structure (with the notation of~\cite{Borzumati:1999qt}),
\begin{align}
  \Op_{15}^q &=  (\bar{s}P_R b) (\bar{q} P_R q)\,, &
  \Op_{16}^q &=  (\bar{s}_\alpha P_R b_\beta) (\bar{q}_\beta P_R q_\alpha)\,, \notag \\
  \Op_{17}^q &=  (\bar{s}P_R b) (\bar{q} P_L q)\,, &
  \Op_{18}^q &=  (\bar{s}_\alpha P_R b_\beta) (\bar{q}_\beta P_L q_\alpha)\,, \notag\\
  \Op_{19}^q &=  (\bar{s} \sigma^{\mu\nu} P_R b) (\bar{q} \sigma_{\mu\nu} P_R q)\,, &
  \Op_{20}^q &=  (\bar{s}_\alpha \sigma^{\mu\nu} P_R b_\beta) (\bar{q}_\beta \sigma_{\mu\nu} P_R q_\alpha)\,.
\end{align}
The primed operators are obtained by interchanging everywhere $P_L \leftrightarrow P_R$. We recall that in the SM only the vector operators receive contributions, while for the scalar/tensor operator the matching contribution is zero. However, NP is expected to contribute to the Hamiltonian also via scalar/tensor operators. We also note that the operators in~\eqref{eqn:Heff_DeltaB1DeltaS1} are redundant since $\Op_1$ and $\Op_2$ can be obtained from $\Op_{3-6}^q$, when $q=c$, via Fierz rearrangements. We will include all NP contributions into the definition of $C_{3-6}^q$ even though for $q=c$ they could be absorbed in $C_1$ and $C_2$ as well. Interestingly, at the leading-logarithmic order only the operators $\Op_{15-20}^q$ mix into the magnetic and chromomagnetic operators $\Op_7$ and $\Op_8$. The vector operators on the other hand mix neither into the magnetic and chromomagnetic nor into the scalar-tensor four-quark operators. The scalar-tensor operators however mix into the vector ones~\cite{Borzumati:1999qt}.

Four fermion operators that involve two right handed currents ($Q_{dd}, Q_{ud}^{(1)}$, and $Q_{ud}^{(8)}$), give the following contribution to the effective Hamiltonian:
\begin{align}
  C_3'^{\, q = d, s, b} &=   -\frac{v^2}{6\Lambda^2}   \tilde{C}_{dd}^{1123, \, 2223, \, 2333}\,, &
  C_5'^{\, q = d, s, b} &=    \frac{v^2}{24 \Lambda^2}   \tilde{C}_{dd}^{1123,\, 2223,\, 2333}\,.
\end{align}
Through a Fierz rearrangement also the operator $\tilde{Q}_{dd}^{1321}$ contributes to
\begin{align}
  C_3'^d &=   -\frac{v^2}{6\Lambda^2}  \frac{1}{N_c}  \tilde{C}_{dd}^{1321}\,, &
  C_5'^d &=   \frac{v^2}{24\Lambda^2}  \frac{1}{N_c}  \tilde{C}_{dd}^{1321}\,, \\
  C_4'^d &=   -\frac{v^2}{3\Lambda^2}  \tilde{C}_{dd}^{1321}\,, &
  C_6'^d &=   \frac{v^2}{12\Lambda^2}  \tilde{C}_{dd}^{1321}\,.
\end{align}
Operators with up-type quarks give:
\begin{align}
   C_3'^{\, q = u,c} &=   -\frac{v^2}{6\Lambda^2}   \tilde{C}_{ud}^{(1)\,  1123, \, 2223}\,, &
  C_5'^{\, q = u, c} &=   \frac{v^2}{24 \Lambda^2}   \tilde{C}_{ud}^{(1)\,  1123,\, 2223}\,, \\
  C_4^{' \, q=u,c} &=  -\frac{v^2}{6\Lambda^2} \tilde{C}^{(8) \, 1123, \, 2223}_{ud}\,,&
  C_6^{' \,q=u,c} &= \frac{v^2}{24 \Lambda^2} \tilde{C}^{(8) \, 1123, \, 2223}_{ud}\,.
\end{align}
In the set $(\bar{L}L)(\bar{R}R)$ in table~\ref{tab:operators1}, the operators with right-handed up-type quarks give the
following contributions:
\begin{align}
  C_3^{\, q = u,c} &=   \frac{2v^2}{3\Lambda^2}   \tilde{C}_{qu}^{(1) \, 2311, \, 2322}\,, &
  C_5^{\, q = u, c} &=   -\frac{v^2}{24 \Lambda^2}   \tilde{C}_{qu}^{(1)\, 2311,\, 2322}\,, \\
  C_4^{\, q = u,c} &=   \frac{2v^2}{3\Lambda^2}   \tilde{C}_{qu}^{(8) \, 2311, \, 2322}\,, &
  C_6^{\, q = u, c} &=   -\frac{v^2}{24 \Lambda^2}   \tilde{C}_{qu}^{(8)\, 2311,\, 2322}\,.
\end{align}
For the same operator set, but with left-handed up-type quarks, we obtain
\begin{align}
  C_3'^{\, q = u,c} &=   \frac{2v^2}{3\Lambda^2}   \check{C}_{qd}^{(1) \, 1123, \, 2223}\,, &
  C_5'^{\, q = u, c} &=   -\frac{v^2}{24 \Lambda^2}   \check{C}_{qd}^{(1)\, 1123, \, 2223}\,,\\
  C_4'^{\, q = u,c} &=   \frac{2v^2}{3\Lambda^2}   \check{C}_{qd}^{(8) \, 1123, \, 2223}\,, &
  C_6'^{\, q = u, c} &=   -\frac{v^2}{24 \Lambda^2}   \check{C}_{qd}^{(8)\, 1123, \, 2223}\,,
\end{align}
where $\check{C}^{(1,8)\, ijkl}_{qd} = V_{im} V^*_{jn} \tilde{C}^{(1,8)\, mnkl}_{qd} $, as defined in section~\ref{sec:operators}. The operators with four down-type quarks give
\begin{align}
  C_3'^{\, q = d, s, b} &=   \frac{2v^2}{3\Lambda^2}   \tilde{C}_{qd}^{(1) \, 1123, \, 2223,\, 3323}\,, &
  C_5'^{\, q = d, s, b} &=   -\frac{v^2}{24 \Lambda^2}   \tilde{C}_{qd}^{(1)\,  1123, \, 2223,\, 3323}\,,\\
  C_3^{\, q = d, s, b} &=   \frac{2v^2}{3\Lambda^2}   \tilde{C}_{qd}^{(1) \, 2311, \, 2322,\, 2333}\,, &
  C_5^{\, q = d, s, b} &=   -\frac{v^2}{24 \Lambda^2}   \tilde{C}_{qd}^{(1)\,  2311, \, 2322,\, 2333}\,,\\
  C_4'^{\, q = d, s, b} &=   \frac{2v^2}{3\Lambda^2}   \tilde{C}_{qd}^{(8) \, 1123, \, 2223,\, 3323}\,, &
  C_6'^{\, q = d, s, b} &=   -\frac{v^2}{24 \Lambda^2}   \tilde{C}_{qd}^{(8)\,  1123, \, 2223,\, 3323}\,,\\
  C_4^{\, q = d, s, b} &=   \frac{2v^2}{3\Lambda^2}   \tilde{C}_{qd}^{(8) \, 2311, \, 2322,\, 2333}\,, &
  C_6^{\, q = d, s, b} &=   -\frac{v^2}{24 \Lambda^2}   \tilde{C}_{qd}^{(8)\,  2311, \, 2322,\, 2333} \,.
\end{align}
Let us now investigate the set of four-fermion operators with the Dirac structure $(\bar{L}L)(\bar{L}L)$. Recalling that for this class of operators we consider only those that fulfill $[ij] \le [kl]$. We obtain the following matching contribution from the vertices involving four left-handed down-type quarks:
\begin{align}
  C_3^{\, q =  s, b} &=   -\frac{v^2}{6\Lambda^2}
  \left[  \tilde{C}_{qq}^{(1) \, 2223, \, 2333}  +\tilde{C}_{qq}^{(3) \,  2223, \, 2333}  \right]\,, \\
  C_5^{\, q =  s, b} &=   +\frac{v^2}{24 \Lambda^2}
  \left[  \tilde{C}_{qq}^{(1) \,  2223, \, 2333}  +\tilde{C}_{qq}^{(3) \,  2223, \, 2333}  \right]\,,\\
  C_3^d &=  -\frac{v^2}{6\Lambda^2}
  \left[  \tilde{C}_{qq}^{(1) \,1123}  +\tilde{C}_{qq}^{(3) \,  1123}+  \frac{1}{N_c}  \left( \tilde{C}_{qq}^{(1)\, 1321}+  \tilde{C}_{qq}^{(3)\, 1321}\right)  \right]\,, \\
  C_5^{d} &=
  +\frac{v^2}{24 \Lambda^2}
	\left[  \tilde{C}_{qq}^{(1) \,1123}  +\tilde{C}_{qq}^{(3) \,  1123}+
  \frac{1}{N_c}  \left( \tilde{C}_{qq}^{(1)\, 1321}+  \tilde{C}_{qq}^{(3)\, 1321}\right)  \right]\,, \\
  C_4^d &=
  -\frac{v^2}{3\Lambda^2}  \left( \tilde{C}_{qq}^{(1)\, 1321}+  \tilde{C}_{qq}^{(3)\, 1321}\right)\,,\\
	C_6^d &=
  +\frac{v^2}{12\Lambda^2}  \left( \tilde{C}_{qq}^{(1)\, 1321}+  \tilde{C}_{qq}^{(3)\, 1321}\right)\,.
\end{align}
From the operators with two left-handed up-type quarks we obtain
\begin{align}
  C_3^{q=u,c} &= -\frac{v^2}{6 \Lambda^2}
  \left( \chi^{(1)}_{u,c} - \chi^{(3)}_{u,c}+\frac{2}{N_c} \Xi^{(3)}_{u,c} \right)\,, \\
  C_5^{q=u,c} &= +\frac{v^2}{24 \Lambda^2}
  \left( \chi^{(1)}_{u,c} - \chi^{(3)}_{u,c} +\frac{2}{N_c}  \Xi^{(3)}_{u,c} \right)\,, \\
  C_4^{q=u,c}&= -\frac{2v^2}{3\Lambda^2}
   \Xi^{(3)}_{u,c} \,, \\
  C_6^{q=u,c}&=+\frac{v^2}{6\Lambda^2}
   \Xi^{(3)}_{u,c} \,,
\end{align}
where the symbols $\chi_q$ and $\Xi_q$ stand for
\begin{align}
  \chi^{(1)}_q &=
  \sum_{[kl]<[23]} \tilde{C}^{(1)\, kl23}_{qq} V_{qk} V_{ql}^*+  \sum_{[kl]>[23]} \tilde{C}^{(1)\, 23kl}_{qq}
   V_{qk} V_{ql}^*   +2 \tilde{C}^{(1)\, 2323}_{qq}V_{qs}V_{qb}^*\,,\\
  \chi^{(3)}_q &=
  \sum_{[kl]<[23]} \tilde{C}^{(3)\, kl23}_{qq} V_{qk} V_{ql}^*+  \sum_{[kl]>[23]} \tilde{C}^{(3)\, 23kl}_{qq}
   V_{qk} V_{ql}^*   +2 \tilde{C}^{(3)\, 2323}_{qq}V_{qs}V_{qb}^*\,,\\
   \Xi^{(3)}_q &=
  \sum_{[2j]<[k3]} \tilde{C}^{(3)\, 2jk3}_{qq} V_{qj}^* V_{qk}  +\sum_{[j3]<[3k]} \tilde{C}^{(3)\, j32k}_{qq} V_{qk}^* V_{qj}  +2 \tilde{C}^{(3)\, 2323}_{qq} V_{qb}^* V_{qs}\,.
\end{align}
Dim-6 operators involving scalar currents generate the following matching contribution for the operators $\Op_{15 - 20}$ in eq.~\eqref{eqn:Heff_DeltaB1DeltaS1} involving $u$ or $c$ quarks:
\begin{align}
  C_{15}^{i=u,c} &=
  \frac{v^2}{2\Lambda^2} \left(    \tilde{C}^{(1)\, ii23}_{quqd}
    -\frac{1}{2N_c} \tilde{C}^{(8)\, ii23}_{quqd}    +\frac{1}{4}  V_{ms}^* V_{in} \tilde{C}^{(8)\, min3}_{quqd}
    \right),\\
  C_{15}^{' \,i=u,c} &=
  \frac{v^2}{2\Lambda^2} \left(    \tilde{C}^{*(1)\, ii32}_{quqd}
    -\frac{1}{2N_c} \tilde{C}^{*(8)\, ii32}_{quqd}    +\frac{1}{4}  V_{in}^* V_{mb} \tilde{C}^{*(8)\, min2}_{quqd}
  \right)\,,\\
  C_{16}^{i=u,c} &=
  \frac{v^2}{4\Lambda^2} \left[
    \tilde{C}^{(8)\, ii23}_{quqd}
    +V_{ms}^* V_{in}
    \left( \tilde{C}^{(1)\, min3}_{quqd}
    -\frac{1}{2N_c} \tilde{C}^{(8)\, min3}_{quqd} \right)
    \right],\\
  C_{16}^{' \,i=u,c} &=
  \frac{v^2}{4\Lambda^2} \left[
    \tilde{C}^{*(8)\, ii32}_{quqd}
    +V_{in}^* V_{mb}
    \left( \tilde{C}^{*(1)\, min2}_{quqd}
    -\frac{1}{2N_c} \tilde{C}^{*(8)\, min2}_{quqd} \right)
  \right],\\
  C_{19}^{i=u,c} &=
  \frac{v^2}{32\Lambda^2}     V_{ms}^* V_{in} \tilde{C}^{(8)\, min3}_{quqd} \,,\\
  C_{19}^{' \,i=u,c} &=
  \frac{v^2}{32\Lambda^2}      V_{in}^* V_{mb} \tilde{C}^{*(8)\, min2}_{quqd}\,,\\
  C_{20}^{i=u,c} &=
  \frac{v^2}{16\Lambda^2}
    V_{ms}^* V_{in} \left(    \tilde{C}^{(1)\, min3}_{quqd}    -\frac{1}{2N_c}\tilde{C}^{(8)\, min3}_{quqd}
    \right)\,,\\
  C_{20}^{' \,i=u,c} &=
  \frac{v^2}{16\Lambda^2}     V_{in}^* V_{mb} \left(    \tilde{C}^{*(1)\, min2}_{quqd}
    -\frac{1}{2N_c}\tilde{C}^{*(8)\, min2}_{quqd}    \right)\,,
\end{align}

The operators $Q_{\varphi q}^{(1)},Q_{\varphi q}^{(3)},Q_{\varphi ud}$ and $Q_{\varphi d}$, involving a $Z$ and $W$ coupling with right-handed fermions, contribute to the four-quark operators in eq.~\eqref{eqn:Heff_DeltaB1DeltaC1} in the following way:
\begin{align}
  C_3^i &=
    \frac{v^2}{\Lambda^2}
    \left[      \frac{1}{3N_c}
    \left( T_3^i+\frac{1}{2} \right)
    \Sigma_{\varphi q}^{i}
    -\left(       \frac{T_3^i}{3}+Q_i\sws
    \right)     \left(
      \tilde C_{\varphi q}^{(1)\,23}
      +\tilde C_{\varphi q}^{(3)\,23}
    \right)  \right]    \,, \\
  C_3^{'\,  i} &= \frac{v^2}{\Lambda^2}
  \left(       \frac{4}{3}T_3^i-Q_i\sws
    \right)\,\tilde C_{\varphi d}^{23} \,, \\
  C_4^{i} &=    \frac{2v^2}{3\Lambda^2}    \left( T_3^i+\frac{1}{2} \right)
  \Sigma_{\varphi q}^{i} \,  ,\\
  C_5^i &= \frac{v^2}{\Lambda^2}  \left[
    \frac{T_3^i}{12}    \left(\tilde C_{\varphi q}^{(1)\,23}+\tilde C_{\varphi q}^{(3)\,23}\right)
    -\frac{1}{12N_c}    \left( T_3^i+\frac{1}{2} \right)    \Sigma_{\varphi q}^{i}
  \right]   \, ,  \\
  C_5^{'\, i} &= - \frac{v^2}{\Lambda^2}\frac{T_3^i}{12}\tilde C_{\varphi d}^{23}  \,, \\
  C_6^{i} &=
    -\frac{v^2}{6\Lambda^2}    \left( T_3^i+\frac{1}{2} \right)
    \Sigma_{\varphi q}^{i}  \, ,\\
  C_{18}^{i} &=    - \frac{v^2}{\Lambda^2} \left( T_3^i+\frac{1}{2} \right)
    V_{is}^* \, \tilde{C}^{i3}_{\varphi ud}\,,\\
  C_{18}^{'\, i} &=    - \frac{v^2}{\Lambda^2} \left( T_3^i+\frac{1}{2} \right)
    V_{ib} \, \tilde{C}^{*i2}_{\varphi ud}\,,
\end{align}
where $i=u,d,c,s,b$ and $Q_i$ and $T_3^i$ denote its charge and third isospin component, respectively.
Moreover we introduced the short notation
$\Sigma^{i}_{\varphi q} =
\tilde C_{\varphi q}^{(3)\, j3} V_{ij} V_{is}^*
+\tilde C_{\varphi q}^{(3)\, 2j} V_{ib} V_{ij}^*$.

The operators involving a vector-current with left-handed quarks directly appear at tree level in the coefficients for $\Op_9,\Op_{10}$ in eq.~\eqref{eqn:O10}:
\begin{align}
  C^{ij}_9 &=
  \frac{\pi}{\alpha}  \frac{v^2}{\Lambda^2}     \left[ \tilde{C}^{(1) \,ij23}_{\ell  q}
  + \tilde{C}^{(3) \,ij23}_{\ell  q}     + \tilde{C}^{23ij}_{qe}  \right]\,, \\
  C^{ij}_{10} &=  \frac{\pi}{\alpha}  \frac{v^2}{\Lambda^2}
  \left[\tilde{C}^{23ij}_{qe}     -\tilde{C}^{(1) \,ij23}_{\ell  q}
  - \tilde{C}^{(3) \,ij23}_{\ell  q}    \right]\,,
  \label{eqn:Wilson9-10}
\end{align}
where the indices $i,j = 1,2,3$, corresponding to $e,\mu$ and $\tau$. Similar contributions appear for the operators $\Op'_9,\Op'_{10}$ from vector-currents involving right-handed quarks:
\begin{align}
  C'^{ij}_9 &=
 \frac{\pi}{\alpha} \frac{v^2}{\Lambda^2}
 \left[   \tilde{C}_{\ell  d}^{ij23} + \tilde{C}_{ed}^{ij23} \right]\,,\\
  C'^{ij}_{10} &=  \frac{\pi}{\alpha}  \frac{v^2}{\Lambda^2}
  \left[     \tilde{C}_{ed}^{ij23}-    \tilde{C}_{\ell  d}^{ij23}   \right]\,.
  \label{eqn:Wilson9-10prime}
\end{align}
Scalar operators contribute to the coefficients of $\Op'_P, \Op'_S$:
\begin{equation}
  C'^{ij}_S = C'^{ij}_P =  \frac{v^2}{4\Lambda^2} \,  \tilde{C}^{ij23}_{\ell edq}\,.
  \label{eqn:WilsonSPprime}
\end{equation}
Also, for the operators $\Op_P, \Op_S$ we have
\begin{equation}
  C^{ij}_S = - C^{ij}_P =   \frac{v^2}{4\Lambda^2} \,   \tilde{C}^{* ji32}_{\ell edq}\,,
  \label{eqn:WilsonSP}
\end{equation}
where the hermitian conjugate of the operator $Q_{\ell edq}^{ijmn}$ is defined as
$\tilde{C}^{* \, ijmn}_{\ell edq} \,
\left( \bar{e}_R^j \ell^i_L \right) \,
\left( \bar{q}^{n}_L d_R^m \right)$\,.
These results agree with those in \cite{Alonso:2014csa} in the case of lepton flavor conservation.
Also the operators $Q_{dB}$ and $Q_{dW}$ appear already at tree-level in the effective Hamiltonian through $\Op_7$ and $\Op_7^{'}$:
\begin{align}
  C_7 &=
  2 \sqrt{2} \sw
  \frac{\pi}{\alpha}
  \frac{M_W}{m_b}
  \frac{v^2}{\Lambda^2}
  \left(   \cw \tilde{C}_{dB}^{23} - \sw\tilde{C}_{dW}^{23}  \right)\,,\\
  C_7^{'} &=
  2 \sqrt{2} \sw
  \frac{\pi}{\alpha}
  \frac{M_W}{m_b}
  \frac{v^2}{\Lambda^2}
  \left(   \cw \tilde{C}_{dB}^{* \, 32} - \sw\tilde{C}_{dW}^{* \, 32}
  \right)\,.
\end{align}
The operators $\Op_9$ and $\Op_{10}$, and similarly $\Op^{\prime}_9$ and $\Op^{\prime}_{10}$, receive the following lepton flavor conserving tree-level contribution through the effective $\bar{s}$-$b$-$Z$ coupling appearing in the operators $Q_{\varphi d}, Q_{\varphi q}^{(1)}$ and $Q_{\varphi q}^{(3)}$:
\begin{align}
  C^{ii}_9 &=  \frac{\pi}{\alpha}
   \frac{v^2}{\Lambda^2}  \left(   \tilde{C}_{\varphi q}^{(1) \, 23}   +\tilde{C}_{\varphi q}^{(3) \, 23}   \right)
  \left( -1+4\sws \right)\,,\\
  C^{ii}_{10} &=\frac{\pi}{\alpha}
  \frac{v^2}{\Lambda^2}    \left(   \tilde{C}_{\varphi q}^{(1) \, 23}   + \tilde{C}_{\varphi q}^{(3) \, 23} \right)\,, \\
  C^{' \, ii}_9 &= \frac{\pi}{\alpha}
  \frac{v^2}{\Lambda^2}  \, \tilde{C}_{\varphi d}^{23}
  \left( -1+4\sws \right)  \,,\\
  C^{' \,ii}_{10} &= \frac{\pi}{\alpha}  \frac{v^2}{\Lambda^2}\,   \tilde{C}_{\varphi d}^{23}\,.
\end{align}
The operator $Q_{dG}$ contributes to the Wilson coefficients of $\mathcal{O}_8$ and $\mathcal{O}_8'$ in the following way:
\begin{align}
  C_8 &=
  \sqrt{2} \, \frac{8 \pi^2}{g \, g_s }
  \frac{M_W}{m_b}
  \frac{v^2}{\Lambda^2}  \tilde{C}_{dG}^{23}\, ,\\
  C^{\prime}_8 &=
  \sqrt{2} \, \frac{8 \pi^2}{g \, g_s }
  \frac{M_W}{m_b}
 \frac{v^2}{\Lambda^2}  \tilde{C}_{dG}^{*\, 32}\,,
\end{align}
where $g$ and $g_s$ are the $SU(2)_L$ and $SU(3)_C$ coupling constants, respectively.
Interestingly, as already noted in ref.~\cite{Alonso:2014csa}, there is no matching contribution to tensor operators at the dimension-six level.

The tree level contribution to the four-quark scalar operators stemming from the operator $Q_{d\varphi}$ is given by
\begin{align}
  C_{15}^b &= C_{17}^b =  -\frac{M_W m_b}{m_h^2} \frac{\sw}{\sqrt{2} e}
  \frac{v^2}{\Lambda^2} \tilde{C}^{23}_{d\varphi}\,, \\
  C_{15}^{'b} &= C_{17}^{'b} =  -\frac{M_W m_b}{m_h^2} \frac{\sw}{\sqrt{2} e}
  \frac{v^2}{\Lambda^2} \tilde{C}^{*\, 32}_{d\varphi}\,.
\end{align}

\section{One-loop matching corrections}
\label{sec:1-loop}

In this section we analyze the leading one-loop matching corrections to the $b \to s$ transitions arising from the dimension-six operators in~\eqref{eqn:Leff}. Let us define what we mean by ``leading'' one-loop matching corrections.
First of all, if one of the gauge invariant operators can contribute already at tree-level to $b\to s$ transitions, a calculation of loop effects is not necessary, since the corresponding Wilson coefficient would already be stringently constrained. Therefore, the loop contribution would only be a subleading effect. With this argument, one can already eliminate all operators that do not contain right-handed up-type quarks: left-handed up quarks always come with their $SU(2)_L$ down quark partner that then contributes to the Hamiltonian at the tree level. Note that it might be possible that an operator containing quark doublets is flavor-violating for up-type quarks but flavor conserving concerning down-type quarks (i.e. not contributing $b\to s$ transitions due to an alignment in flavor space). However, we do not consider this possibility here and focus on operators with up-quark $SU(2)_L$ singlets. Therefore, we are left with the operators given in table~\ref{tab:operators2}.

In the following, we will identify six different classes of matching effects which can be numerically relevant and discuss each of them in a separate subsection. We have the following contributions of gauge invariant dimension-six operators to the ones of the $B$ physics Hamiltonian:

\begin{enumerate}
	\item 4-fermion operators to 4-fermion operators ($\Delta B =\Delta S =1$).
	\item 4-fermion operators to 4-fermion operators ($\Delta B =\Delta S =2$).
	\item 4-fermion operators to $O_7$ and $O_8$.
	\item Right-handed $Z$ couplings to $O_9$, $O_{10}$ and $O^q_{3-6}$.
	\item Right-handed $W$ couplings to $O_7$ and $O_8$.
	\item Magnetic operators to $O_7$, $O_8$, $O_9$, $O_{10}$ and $O_4^q$.
\end{enumerate}

We perform the matching of the operators in table~\ref{tab:operators2} by integrating out the heavy degrees of freedom represented by the Higgs and the top quark, together with the $W$ and $Z$ bosons.
The amplitudes are evaluated at vanishing external momenta, setting all lepton and quark masses to zero except for the top quark mass. To calculate the contribution to the magnetic operators $\Op_7$ and $\Op_8$, as well as the photon and gluon penguins, we expanded the amplitudes up to the second order in external momenta and small quark-masses. In order to check our result we performed the calculation in a general $R_\xi$ gauge, and we explicitly verified the cancellation of the $\xi$ dependent part in the final results.

In several cases, the amplitudes have ultraviolet (UV) divergences. Such divergences signal the running and/or the mixing of different gauge invariant operators between the NP scale $\Lambda$ and the EW scale. The divergences can be (and are) removed via renormalization for which we choose the $\overline{\rm MS}$ scheme. The residual finite terms constitute in these cases the matching result. To indicate the exact origin of the logarithms, we used the notation $\log(m_t^2/\mu_W^2)$ for the one-loop contributions where only the top quark appears in the loop internal legs, while $\log(M_W^2/\mu_W^2)$ signals the presence of at least one $W$-boson in the loop.

\begin{table}[t]
  \centering
  \renewcommand{\arraystretch}{1.2}
  \small
  \begin{tabular}{||c|c|c|c|c|c||}
    \hline \hline
    \multicolumn{2}{||c|}{$\psi^2 X \varphi$} &
    \multicolumn{2}{|c|}{$(\bar{R}R)(\bar{R}R)$} &
    \multicolumn{2}{|c||}{$(\bar{L}L)(\bar{R}R)$} \\ \hline
    $Q_{uW}$ & $( \bar{q}_i \sigma^{\mu\nu} u_j )\, \tau^{\mysmall I} \,\tilde{\varphi}\,W^{\mysmall I}_{\mu\nu}$  &
    $Q_{eu}$  &  $(\bar e_i \gamma_\mu e_j)(\bar{ u}_k \gamma^\mu u_l)$ &
    $Q_{\ell u}$  &  $(\bar \ell_i \gamma_\mu \ell_j)(\bar u_k \gamma^\mu u_l)$ \\
    $Q_{uB}$ &     $( \bar{q}_i \sigma^{\mu\nu} u_j )\,\tilde{\varphi}\, B_{\mu\nu}$ &
    $Q_{uu}$  &   $(\bar{ u}_i \gamma_\mu u_j)(\bar{ u}_k \gamma^\mu u_l)$ &
    $Q_{qu}^{(1)}$  &   $(\bar q_i \gamma_\mu q_j)(\bar u_k \gamma^\mu u_l)$ \\
    $Q_{uG}$ & $(\bar{q}_i \sigma^{\mu\nu} T^A u_j) \tilde{\varphi} G^A_{\mu\nu}$ &
    $Q_{ud}^{(1)}$ & $(\bar{ u}_i \gamma_\mu u_j)(\bar{ d}_k \gamma^\mu d_l)$ &
    $Q_{qu}^{(8)}$  &   $(\bar{ q}_i \gamma_\mu T^A q_j)(\bar u_k \gamma^\mu T^A u_l)$ \\
    \cline{1-2} \cline{5-6}
    \multicolumn{2}{||c|}{$ \psi^2 \varphi^2 D$} &
     $Q_{ud}^{(8)}$ & $(\bar{ u}_i \gamma_\mu T^A u_j)(\bar{ d}_k \gamma^\mu T^A d_l)$ &
     \multicolumn{2}{|c||}{$ (\bar{L}R)(\bar{L}R)$} \\
    \cline{1-2} \cline{5-6}
    $Q_{\varphi ud}$ & $(\tilde{\varphi}^\dagger \, i  D_\mu\varphi )( \bar{u}_i \gamma^\mu d_j )$ &
    & &
    $Q_{quqd}^{(1)}$ &
    $(\bar q^a_i u_j) \eps_{ab} (\bar q^b_k d_l)$ \\
    $Q_{\varphi u}$ & $( \varphi^\dagger i \Dlr_\mu  \varphi )     ( \bar{u}_i\gamma^\mu u_j )$ &
    & &
    $Q_{quqd}^{(8)}$ &
    $(\bar q^a_i T^{\mysmall A} u_j) \eps_{ab} (\bar q^b_k T^{\mysmall A} d_l)$\\
    \hline \hline
  \end{tabular}
  \caption{Dim-6 operators that contribute to $b \to s$ transitions at the one-loop level.}
  \label{tab:operators2}
\end{table}
\subsection[Contribution of 4-fermion operators to 4-fermion operators ($\Delta B =\Delta S =1$)]
{\boldmath Contribution of 4-fermion operators \\to 4-fermion operators ($\Delta B =\Delta S =1$)}
\label{4fo}
\begin{figure}[t]
  \centering
  \subfloat[][]{\includegraphics[width=0.24\textwidth]
    {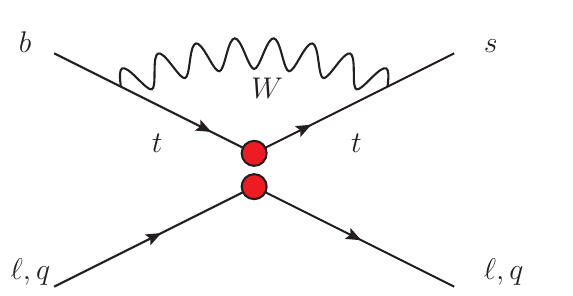} \label{fig:bsmumuqqll}}
  \subfloat[][]{\includegraphics[width=0.24\textwidth]
    {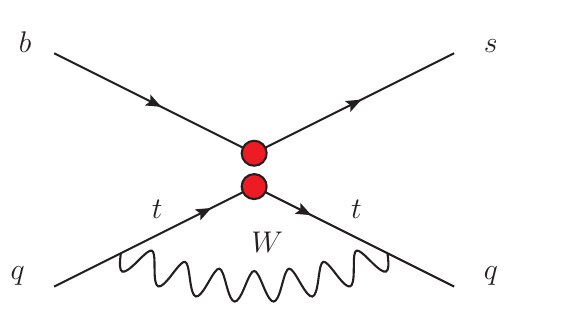} \label{fig:bsmumuqqll2}}
  \subfloat[][]{\includegraphics[width=0.24\textwidth]
    {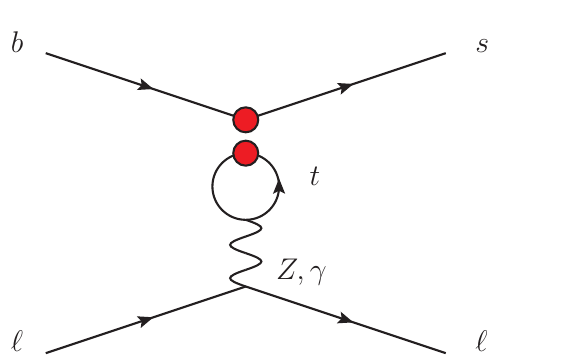} \label{fig:bsmumuQff4q_trace}}
  \subfloat[][]{\includegraphics[width=0.24\textwidth]
    {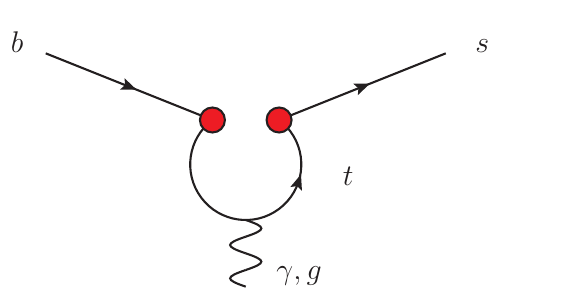} \label{fig:bsmumuQff4q_oneline}}
    \caption{One-loop diagrams in unitary gauge contributing to the low energy theory generated by the four-fermion operators in table~\ref{tab:operators2}. }
  \label{fig:bs2mumuQff}
\end{figure}
We start by reporting the matching contribution to the semi-leptonic operators $\Op_9$ and $\Op_{10}$ from four-fermion operators that couple up-type quarks and charged leptons: $Q_{\ell u }$ and $Q_{eu}$. Obviously, only a charged particle (i.e.\ the $W$ and the charged Goldstone) can give a contribution to a $bs$ operator which is only possible via a genuine vertex correction. Moreover, the result turns out to be proportional to $m_{u_j}^2$. Therefore, we include only the top-quark contribution while $u$ or $c$ quark effects are vanishing in the massless limit.

Calculating the diagram in figure~\ref{fig:bsmumuqqll} (and the analogous Goldstone contribution unless one is working in unitary gauge) gives the following matching contributions:
\begin{align}
  C^{ij}_9 &=   \frac{\lambda_t}{\sws } \,  \frac{v^2}{\Lambda^2} \,  \left[ \tilde{C}_{\ell u}^{ij33} +\tilde{C}_{eu}^{ij33}   \right] I(\xt)\,,\\
  C^{ij}_{10} &=   \frac{\lambda_t}{\sws } \,  \frac{v^2}{\Lambda^2} \,  \left[ \tilde{C}_{eu}^{ij33} - \tilde{C}_{\ell u}^{ij33} \right]   I(\xt)\,,
\end{align}
where $\xt=m_t^2/M_W^2$ and
\begin{equation}
  I(\xt) = \frac{\xt}{16}  \left[  -\ln \left( \frac{M_W^2}{\mu^2_W} \right)    +\frac{\xt-7}{2(1-\xt)}   -\frac{\xt^2-2\xt+4}{(\xt-1)^2}\ln \left(\xt\right)  \right]\,.
  \label{eqn:funI}
\end{equation}
The four-fermion operators involving only quark fields can also contribute to $C^{(\prime)}_9$ and $C^{(\prime)}_{10}$ through a closed top loop (figure~\ref{fig:bsmumuQff4q_trace}) to which an off-shell $Z$ or photon is attached. In this case the contribution is evidently lepton flavor conserving:
\begin{align}
   C^{ii}_9 &=   \tilde{C}_{qu}^{(1) \, 2333} \,   \frac{v^2}{\Lambda^2}    \left( \frac{3\xt}{8 \sws} -\frac{3\xt}{2}-\frac{2}{3} \right)     \ln \left(\frac{m_t^2}{\mu^2_W}\right)\,,\\
   C'^{ii}_9 &=   \tilde{C}_{ud}^{(1) \, 3323} \,   \frac{v^2}{\Lambda^2}   \left( \frac{3\xt}{8 \sws} -\frac{3\xt}{2}-\frac{2}{3} \right)     \ln \left(\frac{m_t^2}{\mu^2_W}\right)\,,\\
   C^{ii}_{10} &=    - \tilde{C}_{qu}^{(1) \, 2333} \,   \frac{v^2}{\Lambda^2}
   \frac{3\xt}{8 \sws}     \ln \left(\frac{m_t^2}{\mu^2_W}\right)\,,\\
   C'^{ii}_{10} &=    -\tilde{C}_{ud}^{(1) \, 3323} \,   \frac{v^2}{\Lambda^2}
   \frac{3\xt}{8 \sws}     \ln \left(\frac{m_t^2}{\mu^2_W}\right)\,.
\end{align}

Furthermore, through a $W$-boson exchange  (figure~\ref{fig:bsmumuqqll2}) the operators under discussion give a one-loop matching contribution to $\Delta B=\Delta S =1$ four-quark operators of the form:
\begin{align}
  C_3^i &=   \tilde{C}_{qu}^{(1) \, 2333} \,   \frac{\alpha}{4 \pi } \,
   \frac{v^2}{\Lambda^2}   \left\lbrace   \ln \left(\frac{m_t^2}{\mu^2_W}\right)
   \left[     Q_i \left(\frac{3\xt}{2}+\frac{2}{3} \right)     +T_3^i\frac{\xt}{2\sws}    \right]
		+ \frac{2}{3}    \left(T_3^i-\frac{1}{2}\right)   \frac{ |V_{ti}|^2 \,I(\xt)}{\sws}
   \right\rbrace\,,\\
   C_3^{'i} &=   \tilde{C}_{ud}^{(1) \, 3323} \,   \frac{\alpha}{4 \pi } \,
   \frac{v^2}{\Lambda^2}   \left\lbrace  \ln \left(\frac{m_t^2}{\mu^2_W}\right)
   \left[     Q_i \left(\frac{3\xt}{2}+\frac{2}{3} \right)     -T_3^i\frac{2\xt}{\sws}    \right]
   -\frac{8}{3}   \left(T_3^i -\frac{1}{2}\right)   \,\frac{ |V_{ti}|^2 \,I(\xt)}{\sws}
   \right\rbrace\,,
\end{align}
\begin{align}
   C_4^i &=   \tilde{C}_{qu}^{(8) \, 2333} \,
   \frac{v^2}{\Lambda^2}   \left\lbrace   \frac{\alpha_s}{24 \pi}   \ln \left( \frac{m_t^2}{\mu^2_W} \right)
   +\frac{\alpha}{6 \pi } \,      \left(T_3^i-\frac{1}{2}\right)
     \, \frac{ |V_{ti}|^2 \,I(\xt)}{ \sws }     \right\rbrace\,,\\
    C_4^{'i} &=
   \tilde{C}_{ud}^{(8) \, 3323} \,   \frac{v^2}{\Lambda^2}   \left\lbrace
   \frac{\alpha_s}{24 \pi}   \ln \left( \frac{m_t^2}{\mu^2_W} \right)   -\frac{2 \alpha}{3 \pi } \,   \left(T_3^i-\frac{1}{2}\right)     \, \frac{ |V_{ti}|^2 \,I(\xt)}{ \sws }     \right\rbrace\,,\\
   C_{5}^i &=
   -\tilde{C}_{qu}^{(1) \, 2333} \,    \frac{\alpha}{32 \pi\sws } \,
   \frac{v^2}{\Lambda^2}    \left[   T_3^i \,\xt    \ln \left(\frac{m_t^2}{\mu^2_W}\right)
   + \frac{4}{3}     \left(T_3^i-\frac{1}{2}\right)     \, |V_{ti}|^2 \,I(\xt)    \right]\,
   ,\\
   C_{5}^{'i} &=
   +\tilde{C}_{ud}^{(1) \, 3323} \,   \frac{\alpha}{32 \pi \sws } \,
   \frac{v^2}{\Lambda^2}    \left[   T_3^i \,\xt    \ln \left(\frac{m_t^2}{\mu^2_W}\right)   + \frac{4}{3}     \left(T_3^i-\frac{1}{2}\right)     \, |V_{ti}|^2 \,I(\xt)    \right]\,,\\
   C_{6}^i &=
   \tilde{C}_{qu}^{(8) \, 2333} \,    \frac{\alpha}{24 \pi\sws } \,   \frac{v^2}{\Lambda^2}    \left(\frac{1}{2}-T_3^i\right)     \, |V_{ti}|^2 \,I(\xt) \, ,\\
     C_{6}^{'i} &=    \tilde{C}_{ud}^{(8) \, 3323} \,   \frac{\alpha}{24 \pi \sws } \,   \frac{v^2}{\Lambda^2}
      \left(T_3^i-\frac{1}{2}\right)     \, |V_{ti}|^2 \,I(\xt)\,,
\end{align}
where $Q_i$ is the charge of the quark, $T_3^i=1/2$ for $q=u,c$ and $T_3^i=-1/2$ for $q=d,s,b$.
Four-fermion operators not containing the flavor violating current $\bar{s}b$ contribute to the four-quarks operators in~\eqref{eqn:Heff_DeltaB1DeltaS1} in the following way:
\begin{align}
  C_3^{i=d,s,b} &=
  \lambda_t   \frac{\alpha}{6 \pi \sws}
  \frac{v^2}{\Lambda^2}    \left[       4 \, \tilde{C}_{ud}^{(1)\, 33ii}    -\tilde{C}_{qu}^{(1)\, ii33}
  \right] I(\xt) \,,\\
  C_3^{i=u,c} &=    \lambda_t   \frac{\alpha}{6 \pi \sws}
  \frac{v^2}{\Lambda^2}    \left[    4 \left( \tilde{C}_{uu}^{ii33}-\frac{1}{N_c} \tilde{C}_{uu}^{i33i} \right)
    -\check{C}_{qu}^{(1)\, ii33}   \right] I(\xt)\,,\\
  C_4^{i=d,s,b} &=   \lambda_t   \frac{\alpha}{6 \pi \sws}
  \frac{v^2}{\Lambda^2}  \left[
    4 \, \tilde{C}_{ud}^{(8)\, 33ii}    -\tilde{C}_{qu}^{(8)\, ii33}
  \right] I(\xt)\,,\\
  C_4^{i=u,c} &=    \lambda_t   \frac{\alpha}{6 \pi \sws}
  \frac{v^2}{\Lambda^2}    \left[
    -8 \tilde{C}_{uu}^{i33i}        -\check{C}_{qu}^{(8)\, ii33}
  \right]I(\xt)\,,\\
   C_5^{i=d,s,b} &=     \lambda_t   \frac{\alpha}{24 \pi \sws}
  \frac{v^2}{\Lambda^2}    \left[
     \tilde{C}_{qu}^{(1)\, ii33} -\tilde{C}_{ud}^{(1)\, 33ii}
   \right]  I(\xt)\,,\\
  C_5^{i=u,c} &=    \lambda_t   \frac{\alpha}{24 \pi \sws}
  \frac{v^2}{\Lambda^2}    \left[
   \check{C}_{qu}^{(1)\, ii33}     -\tilde{C}_{uu}^{ ii33}
   +\frac{1}{N_c} \tilde{C}_{uu}^{i33i}   \right]  I(\xt)\,,\\
  C_6^{i=d,s,b} &=
  \lambda_t   \frac{\alpha}{24 \pi \sws}  \frac{v^2}{\Lambda^2}
  \left[        \tilde{C}_{qu}^{(8)\, ii33}   -\tilde{C}_{ud}^{(8)\, 33ii}     \right]  I(\xt)\,,\\
  C_6^{i=u,c} &=
  \lambda_t   \frac{\alpha}{24 \pi \sws}    \frac{v^2}{\Lambda^2}  \left[
    \check{C}_{qu}^{(8)\, ii33}     +2 \, \tilde{C}_{uu}^{i33i}  \right]I(\xt)\,,
\end{align}
where here we used also the notation introduced in section~\ref{sec:operators}: $\check{C}^{(1,8)\, ijkl}_{qu} = V_{im} V^*_{jn} \tilde{C}^{(1,8)\, mnkl}_{qu} $.

\subsection[Contribution of 4-fermion operators to 4-fermion operators ($\Delta B =\Delta S =2$)]
{\boldmath Contribution of 4-fermion operators \\to 4-fermion operators ($\Delta B =\Delta S =2$)}
The Hamiltonian for $B_s$-$\bar{B}_s$ mixing in eq.~\eqref{eqn:Heff_DeltaF2} gets a one-loop matching contribution through the graph in figure~\ref{fig:bsmumuqqll2}:
\begin{align}
  C_1 &=
  \lambda_t \frac{\alpha}{2 \pi \sws}
  \frac{1}{\Lambda^2}  I(\xt)
  \left[
   \left( -1+ \frac{1}{N_c} \right)  \tilde{C}_{qu}^{(8)\, 2333}
   -2\tilde{C}_{qu}^{(1)\, 2333}
 \right]\,, \label{eqn:1loopBBarC1} \\
  C_4 &=
  \lambda_t \frac{\alpha}{\pi \sws}  \frac{1}{\Lambda^2}  I(\xt)   \,
  \tilde{C}_{ud}^{(8)\, 3323}\,,\\
  C_5 &=
  \lambda_t \frac{\alpha}{\pi \sws}  \frac{1}{\Lambda^2}  I(\xt)   \left[
     2 \tilde{C}_{ud}^{(1)\, 3323}   - \frac{1}{N_c}   \tilde{C}_{ud}^{(8)\, 3323} \right]\,.
\label{eqn:1loopBBarC5}
\end{align}
%

\subsection[ Contributions of 4-fermion operators to $O_7$ and $O_8$]
{\boldmath Contributions of 4-fermion operators to $O_7$ and $O_8$}

Four-fermion operators with scalar currents contribute to the low energy Hamiltonian~\eqref{eqn:Heff_DeltaB1DeltaS1} through the diagram in figure~\ref{fig:bsmumuQff4q_oneline}:
\begin{align}
   C_7 =&
   -\frac{1}{6}   \frac{m_t}{m_b}
   \frac{v^2}{\Lambda^2}   \ln \left( \frac{m_t^2}{\mu^2_W} \right)
   \left[ \tilde{C}_{quqd}^{(1) \,2333}+      C_F \, \tilde{C}_{quqd}^{(8) \,2333}   \right]\,,\\
  C_7' =&
  -\frac{1}{6}   \frac{m_t}{m_b}   \frac{v^2}{\Lambda^2}
  \ln \left( \frac{m_t^2}{\mu^2_W} \right)
  \left[ \tilde{C}_{quqd}^{*(1) \, 3332}+    C_F \, \tilde{C}_{quqd}^{*(8) \, 3332}
    \right]\,,\\
   C_8 =&   -\frac{1}{4}   \frac{m_t}{m_b}   \frac{v^2}{\Lambda^2}   \ln \left( \frac{m_t^2}{\mu^2_W} \right)
   \left[ \tilde{C}_{quqd}^{(1) \,2333}-    \frac{1}{2N_c} \tilde{C}_{quqd}^{(8) \,2333}
   \right]\,,\\
   C_8' =&   -\frac{1}{4}   \frac{m_t}{m_b}   \frac{v^2}{\Lambda^2}   \ln \left( \frac{m_t^2}{\mu^2_W} \right)  \left[       \tilde{C}_{quqd}^{*(1) \, 3332}-      \frac{1}{2N_c} \tilde{C}_{quqd}^{*(8) \, 3332}   \right]\,,
\end{align}
where $C_F = (N_c^2-1)/(2N_c)$.
Note that the contribution to $C_7$ or $C_8$ from 4-fermion operators involving vector currents vanishes (excluding QCD corrections).

\subsection[Contributions of right-handed $Z$ couplings to $O_9$, $O_{10}$ and $O^q_{3-6}$]
{\boldmath Contributions of right-handed $Z$ couplings to $O_9$, $O_{10}$ and $O^q_{3-6}$}
The operator $Q_{\varphi u}$, involving only right-handed up-type quarks, gives through a $Z$-penguin (figure~\ref{fig:QuBZpingu2}) a matching contribution to the $\Delta B =\Delta S =1$ Hamiltonian in eq.~\eqref{eqn:Heff_DeltaB1DeltaS1} of the form:
\begin{align}
  C_3^i &=
  - \lambda_t \frac{\alpha}{\pi \sws} \frac{v^2}{\Lambda^2}
  I(\xt)\,  \tilde{C}^{33}_{\varphi u}
  \left( Q_i \sws+\frac{1}{3}T_3^i \right)\,, \\
  C_5^i &=
  \lambda_t \frac{\alpha}{12 \pi \sws} \frac{v^2}{\Lambda^2}
  I(\xt) \, \tilde{C}^{33}_{\varphi u} \,  T_3^i\,,\\
  C_9^{ii} &=
  \frac{\lambda_t}{\sws} \frac{v^2}{\Lambda^2} \, \tilde{C}_{\varphi u}^{33} \,  I(\xt) \left(  -1+4 \sws\right)\,,\\
  C_{10}^{ii} &=
  \frac{\lambda_t}{\sws} \frac{v^2}{\Lambda^2}  \,  \tilde{C}_{\varphi u}^{33} \,   I(\xt)\,,
\end{align}
where $I (\xt)$ has been defined in eq.~\eqref{eqn:funI}. The possibility to probe the anomalous couplings of the $Z$ boson to top quark with rare meson decays were also studied in~\cite{Brod:2014hsa}.
\subsection[Contributions of right-handed $W$ couplings to $O_7$ and $O_8$]
{\boldmath Contributions of right-handed $W$ couplings to $O_7$ and $O_8$}
The operator $Q_{\varphi ud}$ couples the $W$ boson to right-handed quarks, which induces a non-zero contribution only to the magnetic terms $\Op_7, \Op_8$:
\begin{align}
   C_7 &=   \frac{m_t}{m_b} \,
   \frac{v^2}{\Lambda^2}   E_{\mysmall \varphi ud}^7 (\xt)  \,
   \tilde{C}_{\varphi ud}^{33} \,   V_{ts}^*\,,\label{eqn:C7phiud}\\
   C'_7 &=   \frac{m_t}{m_b} \,
   \frac{v^2}{\Lambda^2}     E_{\mysmall \varphi ud}^7 (\xt)  \,
   \tilde{C}_{\varphi ud}^{* \, 32} \,   V_{tb}\,,\\
   C_8 &=   \frac{m_t}{m_b}
   \frac{v^2}{\Lambda^2}     E_{\mysmall \varphi ud}^8 (\xt)  \,
   \tilde{C}_{\varphi ud}^{ 33} \,   V_{ts}^*\,,\label{eqn:C8phiud}\\
   C_8' &=   \frac{m_t}{m_b}   \frac{v^2}{\Lambda^2}
   E_{\mysmall \varphi ud}^8 (\xt)  \,
   \tilde{C}_{\varphi ud}^{* \, 32} \,   V_{tb}\,,
\end{align}
where the $\xt$-functions, in agreement with~\cite{Grzadkowski:2008mf,Drobnak:2011aa}, are
\begin{align}
   E_{\mysmall \varphi ud}^7 (\xt) &=   \frac{-5\xt^2+31\xt-20}{24(\xt-1)^2}   +\frac{ \xt (2-3 \xt)}{4(\xt-1)^3} \ln \left(\xt \right)\,,\\
   E_{\mysmall \varphi ud}^8 (\xt) &=   - \frac{\xt^2+\xt+4}{8(\xt-1)^2}
   +\frac{3 \xt}{4(\xt-1)^3}\ln\left( \xt\right)\,.
\end{align}

\subsection[Contributions of magnetic operators to
  $O_7, O_8, O_9, O_{10}$ and $O_4^q$]
{\boldmath Contributions of magnetic operators to
  $O_7, O_8, O_9, O_{10}$ and $O_4^q$}
%
\begin{figure}[t]
  \centering
  \subfloat[][]{\includegraphics[width=0.24\textwidth]
    {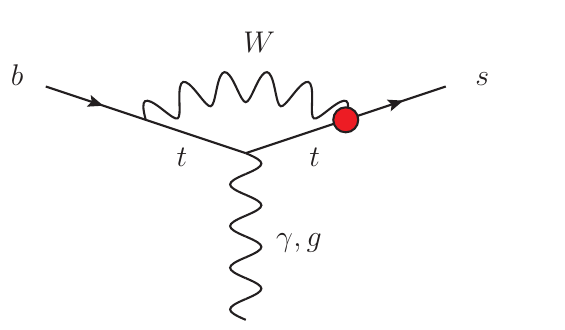} \label{fig:QuBW}}
  \subfloat[][]{\includegraphics[width=0.24\textwidth]
    {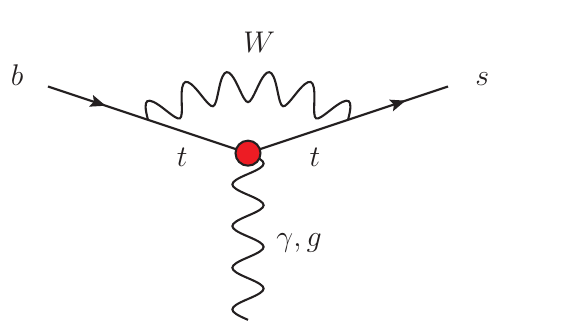} \label{fig:QuBW2}}
  \subfloat[][]{\includegraphics[width=0.24\textwidth]
    {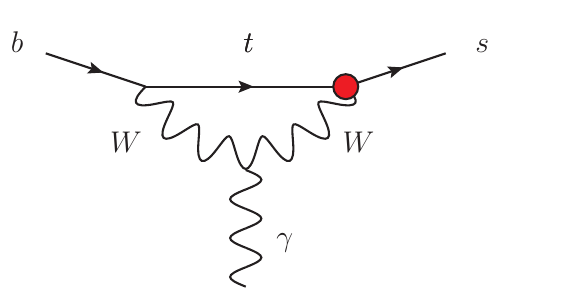} \label{fig:QuBWW}}
  \subfloat[][]{\includegraphics[width=0.24\textwidth]
    {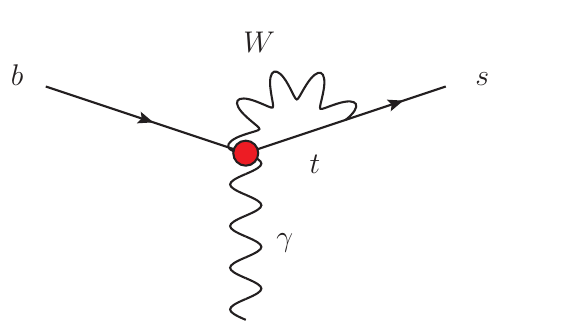} \label{fig:QuBWZ}}\\
  \subfloat[][]{\includegraphics[width=0.24\textwidth]
    {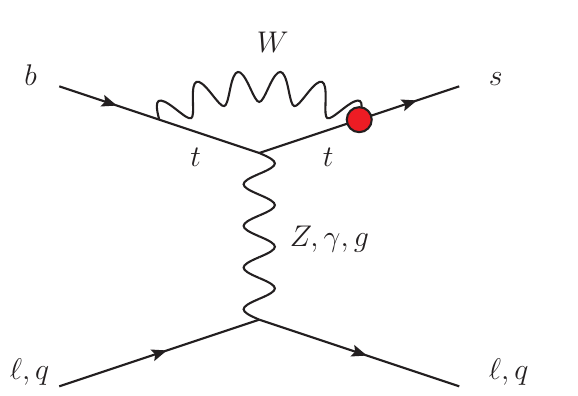} }
  \subfloat[][]{\includegraphics[width=0.24\textwidth]
    {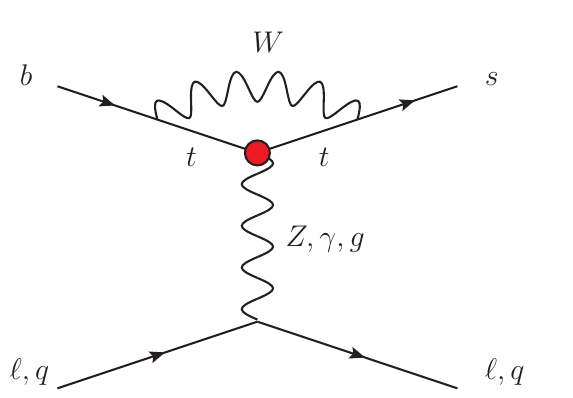} \label{fig:QuBZpingu2} }
  \subfloat[][]{\includegraphics[width=0.24\textwidth]
    {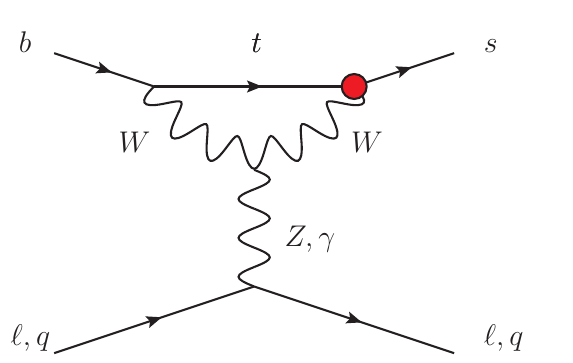} }
  \subfloat[][]{\includegraphics[width=0.24\textwidth]
    {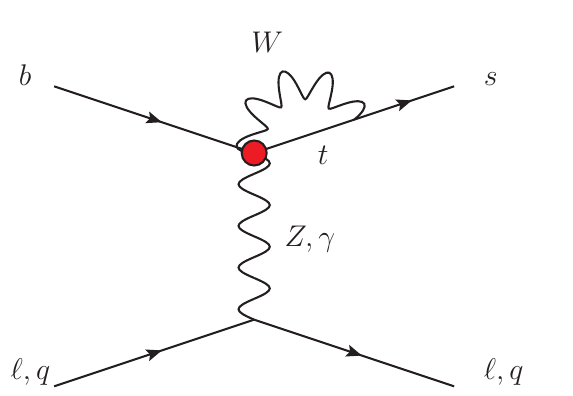} }
    \caption{\small One-loop diagrams in the unitary gauge for $b \to s V$ transitions (with $V=Z,\gamma,g$) originating from the operators $Q_{uB}, Q_{uW}$ and $Q_{uG}$.  The red dots represent an operator insertion. For each of these diagrams a symmetric one must also be considered, with the effective operator in the $W$-$t$-$b$ vertex. Box diagrams and self energies on the external legs (not depicted here) must also be included.}
  \label{fig:bs2mumuQub}
\end{figure}

In this subsection we summarize the matching contributions arising from the magnetic operators in table~\ref{tab:operators2}. The operators $Q_{uB}$ and $Q_{uW}$ contribute to the effective Hamiltonian for $b \to s \gamma$ and $b \to s \bar{\ell} \ell$  transitions via the one-loop diagrams in figure~\ref{fig:bs2mumuQub}.

For simplicity, let us first consider the operators $\tilde{C}^{33}_{uW}$ and $\tilde{C}^{33}_{uB}$ that generate an extra term for the top anomalous magnetic moment resulting in a chirality flipping vertex with the $W$ boson.
We will later analyse the case when the vertices with the photon and the $Z$ are flavor violating.
Here we include only the contributions to four-quark operators arising from gluon-penguin diagrams, which are of $O(\alpha_s)$, and we neglect the subleading EW penguin diagrams, of $O(\alpha)$. We obtained the following contributions to the effective Hamiltonian in eq.~\eqref{eqn:Heff_DeltaB1DeltaS1}:
\begin{align}
  C_4^i &=   \lambda_t   \frac{\alpha_s}{\pi}    \frac{m_t}{M_W}  \frac{\sqrt{2}\, v^2}{\Lambda^2} \,
  A_{\mysmall uW}(\xt) \,   \Re ( \tilde{C}^{33}_{uW})\,, \\
  C_7 &=
  \lambda_t   \frac{m_t}{M_W}   \frac{\sqrt{2}\, v^2}{\Lambda^2}
  \left\lbrace   \tilde{C}^{33}_{uW} E^7_{\mysmall uW} (\xt)
  +\tilde{C}^{*33}_{uW} \, F^7_{\mysmall uW} (\xt)
  +\frac{\cw}{\sw}   \left[    \tilde{C}^{33}_{uB} E^7_{\mysmall uB} (\xt)
  +\tilde{C}^{*33}_{uB} \, F^7_{\mysmall uB} (\xt)
  \right]  \right\rbrace\,,\\
  C_8 &=
  \lambda_t   \frac{m_t}{M_W}
  \frac{\sqrt{2}\, v^2}{\Lambda^2} \,
  \left[     \tilde{C}^{33}_{uW}    \, E_{\mysmall uW}^8 (\xt)
  + \tilde{C}^{*33}_{uW}  \, F_{\mysmall uW}^8 (\xt)
  \right]\,,\\
  C_9^{ii} &=
 \lambda_t    \frac{m_t}{M_W}
 \frac{\sqrt{2}\,  v^2}{\Lambda^2}
 \left[     \Re (\tilde{C}^{33}_{uW} )
   \left(  \frac{Y_{uW}(\xt)}{\sws}  -Z_{uW}(\xt)  \right)
   -\frac{\cw}{\sw}  \Re (\tilde{C}^{33}_{uB})  Z_{uB}(\xt)
  \right]\,, \\
  C_{10}^{ii} &=
  -  \, \lambda_t  \frac{m_t}{M_W}
  \frac{\sqrt{2}\,v^2}{\Lambda^2}
  \frac{Y_{\mysmall uW}(\xt)}{\sws}  \Re (\tilde{C}^{33}_{uW} )\,,
\end{align}
where the explicit expressions for the $\xt$-dependent functions are
\begin{align}
  E^7_{\mysmall uW} (\xt) &=
  \frac{1}{8} \ln \left(\frac{M_W^2}{\mu^2_W} \right)
  +\frac{-9 \xt^3+63\xt^2-61\xt+19}{48(\xt-1)^3}
  +\frac{3\xt^4-12\xt^3-9\xt^2+20\xt-8}{24(\xt-1)^4} \ln \left(\xt\right),\\
  F^7_{\mysmall uW} (\xt) &=
  -\frac{3\xt^3-17\xt^2+4\xt+4}{24(\xt-1)^3}
  +\frac{\xt(2-3\xt)}{4(\xt-1)^4} \ln \left(\xt\right)\,,\\
  E^7_{\mysmall uB} (\xt) &=
  -\frac{1}{8} \ln \left(\frac{M_W^2}{\mu^2_W}\right)
  -\frac{(\xt+1)^2}{16(\xt-1)^2}
  -\frac{\xt^2(\xt-3)}{8(\xt-1)^3}  \ln \left(\xt\right)\,,\\
  F^7_{\mysmall uB} (\xt) &=
  -\frac{1}{8}\,,\\
  E_{\mysmall uW}^8 (\xt) &=
  \frac{3\xt^2-13\xt+4}{8(\xt-1)^3}
  +\frac{5\xt-2}{4(\xt-1)^4} \ln \left(\xt\right)\,, \\
  F_{\mysmall uW}^8 (\xt) &=
  \frac{\xt^2-5\xt-2}{8(\xt-1)^3}
  +\frac{3\xt}{4(\xt-1)^4} \ln \left(\xt\right)\,, \\
  A_{\mysmall uW}(\xt) &=
  \frac{5\xt^2-19\xt+20}{24(\xt-1)^3}
  +\frac{\xt-2}{4(\xt-1)^4} \ln \left(\xt\right)\,,\\
  Y_{\mysmall uW} (\xt) &=
  \frac{3\xt}{4(\xt-1)}
  -\frac{3 \xt}{4(\xt-1)^2}\ln \left(\xt\right)\,,\\
  Z_{\mysmall uW}(\xt) &=
     \frac{99\xt^3-136\xt^2-25\xt+50}{36(\xt-1)^3}
     -\frac{24\xt^3-45\xt^2+17\xt+2}{6(\xt-1)^4}\ln \left(\xt\right)\,,\\
  Z_{\mysmall uB} (\xt) &=
  -\frac{\xt^2+3\xt-2}{4(\xt-1)^2}  +\frac{3\xt-2}{2(\xt-1)^3}\ln \left(\xt\right)\,.
\end{align}
We found that the expressions for the functions $E_{\mysmall uW}^i$,$F_{\mysmall uW}^i$,$Y_{\mysmall uW}$ and $Z_{\mysmall uW}$ are in agreement with the results reported in~\cite{Grzadkowski:2008mf,Drobnak:2011aa}, while $A_{uW}, Z_{uB}, E_{\mysmall uB}^7$ and $F_{\mysmall uB}^7$ are new to the best of our knowledge. Note that the effect on the magnetic operators $\Op_7$ and $\Op_8$ is divergent while it is finite for the four-fermion operators. Moreover, all these effects scale like $1/\Lambda^2$ and do not possess an additional suppression by $1/M_W^2$.

Now we turn our attention to the operators $Q_{uW}^{i3}$ and $Q_{uB}^{i3}$, where $i=1,2$.\footnote{The effect of a right-handed $W$-$t$-$d$ coupling on $b\to d\gamma$ was studied in ref.~\cite{Crivellin:2011ba}.}
These operators lead to an anomalous $W$-$t$-$d^i$ coupling, plus two flavor-violating neutral currents $(Z/\gamma) tc$ and $(Z/\gamma) tu$, so then in the diagram~\ref{fig:QuBW2} one top quark propagator becomes $q=u,c$.
However, we recall that this amplitude is non-zero only for the $\gamma$ penguin, or the transition $b \to s \gamma$ --- the effective coupling is proportional to $\sigma^{\mu\nu}q_\nu$, where $q$ is the momentum of the boson. Only the functions arising from a $\gamma$ penguin will be modified in this case, i.e.\ the functions $Z,E^7,F^7$. Repeating the calculations performed for $\tilde{C}^{33}_{uB}$ and $\tilde{C}_{uW}^{33}$ we obtain the following results for the matching:
\begin{align}
  C_4^i &=
  \frac{\alpha_s}{\pi}   \frac{m_t}{M_W}    \frac{\sqrt{2}\, v^2}{\Lambda^2} \,    A_{\mysmall uW}(\xt) \,   \Sigma_{uW}\,,\\
  C_{7} &=
  \frac{m_t}{M_W}  \frac{\sqrt{2}\, v^2}{\Lambda^2}
  \left\lbrace      \tilde{C}^{i3}_{uW} V_{is}^* V_{tb} E^{'7}_{\mysmall uW} (\xt)
    + \tilde{C}^{*\, i3}_{uW} V_{ib} V^*_{ts} F^{7}_{\mysmall uW} (\xt)
    \right. \notag \\
   & \left.
   + \frac{\cw}{\sw}   \left[
     \tilde{C}^{i3}_{uB} V_{is}^* V_{tb} E^{'7}_{\mysmall uB} (\xt)
     + \tilde{C}^{*\, i3}_{uB} V_{ib} V^*_{ts} F^{7}_{\mysmall uB} (\xt)
   \right]
  \right\rbrace\,, \\
 C_8 &=
  \frac{m_t}{M_W}  \frac{\sqrt{2}\, v^2}{\Lambda^2} \,
  \left[    \tilde{C}^{i3}_{uW} V_{is}^* V_{tb}    \, E_{\mysmall uW}^8 (\xt)
    + \tilde{C}^{*\, i3}_{uW} V_{ib} V^*_{ts} \, F_{\mysmall uW}^8 (\xt)  \right] \,, \\
    C_9^{ii} &=
  \frac{m_t}{M_W}  \frac{\sqrt{2}\,  v^2}{\Lambda^2}
  \left[      \Sigma_{uW}   \left(  \frac{Y_{uW}(\xt)}{\sws}  -Z_{uW}'(\xt)  \right)
    -\frac{\cw}{\sw}  \Sigma_{uB}  Z_{uB}'(\xt)
  \right]\,, \\
  C_{10}^{ii} &=
  - \frac{m_t}{M_W}  \frac{\sqrt{2}\,v^2}{\Lambda^2}
  \frac{Y_{\mysmall uW}(\xt)}{\sws}  \Sigma_{uW}\,,
\end{align}
where
$\Sigma_{uW} =  ( \tilde{C}^{i3}_{uW} V_{is}^* V_{tb}  + \tilde{C}^{*\, i3}_{uW} V_{ib} V^*_{ts} )/2$
and
$\Sigma_{uB} =  ( \tilde{C}^{i3}_{uB} V_{is}^* V_{tb}  + \tilde{C}^{*\, i3}_{uB} V_{ib} V^*_{ts} )/2$ (the summation over $i=1,2$ is implied). The new functions introduced above are:
\begin{align}
  Z_{uW}'(\xt) &=
  \frac{54 x_t^3-59 x_t^2-35 x_t+34}{18 \left(x_t-1\right){}^3}
  -\frac{15 x_t^3-27 x_t^2+10 x_t+1}{3 \left(x_t-1\right){}^4} \ln \left(\xt\right),\\
  Z_{uB}'(\xt) &=  \frac{1}{1-\xt}\ln \left(\xt\right)\,,\\
  E^{'7}_{\mysmall uW} (\xt) &=
  \frac{1}{8}\ln \left(\frac{M_W^2}{\mu^2_W} \right)
  +\frac{-3 \xt^3+63\xt^2-67\xt+19}{48(\xt-1)^3}
  +\frac{3\xt^4-18\xt^3-3\xt^2+20\xt-8}{24(\xt-1)^4} \ln \left(\xt\right)\,,\\
  E^{'7}_{\mysmall uB} (\xt) &=
  -\frac{1}{8} \ln \left(\frac{M_W^2}{\mu^2_W}\right)
  +\frac{\xt+1}{16(\xt-1)}   -\frac{\xt^2}{8(\xt-1)^2}  \ln \left(\xt\right)\,.
\end{align}

The operator $Q_{uG}^{33}$ gives a chromo-magnetic coupling with the top quark, that contributes at one-loop to $\Op_8$ and $\Op_4$ through the gluon-penguin diagrams in figure~\ref{fig:QuBW2},\ref{fig:QuBZpingu2}. The explicit matching contributions are
\begin{align}
  C_4^i &=
  \lambda_t  \frac{g g_s}{16 \pi^2}    \frac{m_t}{M_W} \frac{\sqrt{2} \,v^2}{\Lambda^2}    \Re (\tilde{C}_{uG}^{33})  A_{uG}(\xt)\,, \\
  C_8 &=  \lambda_t \frac{g}{g_s}    \frac{m_t}{M_W} \frac{\sqrt{2} \, v^2}{\Lambda^2}
  \left[  \tilde{C}_{uG}^{33} \, E^8_{uG}(\xt)    +\tilde{C}_{uG}^{*33} \, F^8_{uG}(\xt)  \right]\,,
\end{align}
where $A_{uG} = Z_{uB}$, $ E^8_{uG} = E^7_{uB}$ and $ F^8_{uG} = F^7_{uB}$.
Moreover, the operators $Q_{uG}^{i3}$ lead to a flavor violating neutral current involving a gluon and up-type quarks, whose effects in the effective Hamiltonian are
\begin{align}
  C_4^i &=   \frac{g g_s}{16 \pi^2}
  \frac{m_t}{M_W} \frac{\sqrt{2} \,v^2}{\Lambda^2}
  A'_{uG}(\xt)\,
  \frac{
    \tilde{C}_{uG}^{i3} V_{tb} V_{is}^*
    +\tilde{C}_{uG}^{*i3} V_{ib} V_{ts}^*}{2}\,, \\
  C_8 &=
   \frac{g}{g_s}
  \frac{m_t}{M_W} \frac{\sqrt{2} \, v^2}{\Lambda^2}
  \left[
    \tilde{C}_{uG}^{i3} V_{tb} V_{is}^* \, E^{'8}_{uG}(\xt)
    +\tilde{C}_{uG}^{*i3} V_{ib} V_{ts}^* \, F^8_{uG}(\xt)
  \right]\,,
\end{align}
where $A'_{uG} = Z'_{uB}$ and $E^{'8}_{uG} = E^{'7}_{uB}$.

\section{Phenomenological example}
\label{sec:example}
As an example of phenomenological applications of the matching conditions reported in sections~\ref{sec:tree} and~\ref{sec:1-loop}, we will consider  the operator~$\tilde{Q}_{\varphi ud}^{33}$ that gives rise to a one-loop contribution to $C_7$ and $C_8$ (see eqs.\eqref{eqn:C7phiud} and \eqref{eqn:C8phiud}).
We can employ the inclusive $\bar{B} \to X_s \gamma$ branching ratio to constrain the the Wilson coefficient~$\tilde{C}_{\varphi ud}^{33}$.
Let us denote the Wilson coefficients in~\eqref{eqn:Heff_DeltaB1DeltaS1} as $C_i(\mu) = C_i^{\rm SM}(\mu) + \Delta C_i(\mu)$, where $\Delta C_i(\mu)$ are possible non-SM terms.
The calculation of the contribution to the decay $\bar{B} \to X_s \gamma$ proceeds precisely as in the SM case:
\begin{itemize}
  \item The evolution of the Wilson coefficients in~\eqref{eqn:Heff_DeltaB1DeltaS1}, from the mass scale $\mu=\mu_W$ down to $\mu=\mu_b$, where $\mu_b$ is of the order or $m_b$, by solving the appropriate RGE.
  \item The evaluation of the corrections to the matrix elements $\bra{s\gamma} O_i(\mu)\ket{b}$ at the scale $\mu=\mu_b$, and the subsequent shift induced in the branching ratio $\mathcal{B}(\bar{B} \to X_s \gamma)$.
\end{itemize}
The beyond-SM effect on $\mathcal{B}(\bar{B} \to X_s \gamma)$, driven by the new additive contribution involving $\Delta C_{7,8}$, can be compactly written as
\begin{equation}
  \mathcal{B}^{\rm th} (\bar{B} \to X_s \gamma) \times 10^4 =
  (3.36 \pm 0.23)
  - \frac{ 8.22 \, \Delta C_7+1.99 \, \Delta C_8}{V_{tb}V_{ts}^*},
  \label{eqn:Btheo}
\end{equation}
where $\Delta C_{7,8}$ are defined at the mass scale $\mu_W=160$ GeV~\cite{Misiak:2015xwa}.
The theoretical prediction~\eqref{eqn:Btheo} incorporates NNLO QCD corrections as well as nonperturbative effects.
The input parameters and their uncertainties can be found in Appendix D of ref.~\cite{Czakon:2015exa}.
Moreover it is assumed that the quadratic terms in $\Delta C_{7,8}$ are negligible.

For the purpose of this example we assume $\tilde{C}_{\varphi ud}^{33}$ to be real and we neglect the imaginary part of $V_{tb}$ and $V_{ts}$.
Identifying the non-SM terms $\Delta C_{7,8}$ with the results in eqs.\eqref{eqn:C7phiud} and \eqref{eqn:C8phiud}, and taking into account the current world average~\cite{Amhis:2014hma}
\begin{equation}
  \mathcal{B}^{\rm exp} (\bar{B} \to X_s \gamma) =
  (3.43 \pm 0.21 \pm 0.07) \times 10^{-4},
  \label{eqn:Bexp}
\end{equation}
we can find the current 95\% C.L.\ bounds
\begin{equation}
  -3.3 \times 10^{-3} \le
  \tilde{C}_{\varphi ud}^{33} \left[\mu_W =160 \text{GeV}\right] \frac{v^2}{\Lambda^2} \le
  2.7 \times 10^{-3}.
  \label{eqn:boundbtos}
\end{equation}
This quite strong bound takes place due to a relative enhancement $m_t/m_b$ compared to the SM case: the SM chiral suppression factor $m_b/M_W$ is replaced by the factor $m_t/M_W$~\cite{Cho:1993zb}. It can be interesting to compare~\eqref{eqn:boundbtos} with the $Wtb$ vertex structure searches at the LHC.
The 8 TeV data on the single top quark production cross section and the measurements of the $W$-boson helicity fractions allowed the authors of~\cite{Bernardo:2014vha} to set a bound on $\tilde{C}_{\varphi ud}^{33}\, (v^2/\Lambda^2)$ at the level of $10^{-1}$.
Also, ATLAS searches for anomalous couplings in the $Wtb$ vertex from the measurement of double differential angular decay rates of single top quarks produced in the $t$-channel show similar sensitivities~\cite{Aad:2015yem}.

\section{Conclusions}
\label{sec:conclusions}
In this article, we calculated (at the EW scale) the matching of the gauge invariant dimension-six
 operators on the $B$ physics Hamiltonian (including lepton flavor violating operators) integrating out the top, $W$, $Z$ and the Higgs.
 After performing the EW symmetry breaking and diagonalizing the mass matrices, we first presented the complete tree-level matching coefficients for $b\to s$ and $b\to c$ transitions. Operators involving top quarks do not contribute to $b\to s$ processes at the tree level, as the top is not a dynamical degree of freedom of the $B$ physics Hamiltonian. Therefore, we identified all operators involving right-handed top quarks which can give numerically important contributions at the one loop-level:
\begin{enumerate}
  \item 4-fermion operators to 4-fermion operators ($\Delta B =\Delta S =1$).
  \item 4-fermion operators to 4-fermion operators ($\Delta B =\Delta S =2$).
  \item 4-fermion operators to $O_7$ and $O_8$.
  \item Right-handed $Z$ couplings to $O_9$, $O_{10}$ and $O^q_{3-6}$.
  \item Right-handed $W$ couplings to $O_7$ and $O_8$.
  \item Magnetic operators to $O_7$, $O_8$, $O_9$, $O_{10}$ and $O_4^q$.
\end{enumerate}
Once the necessary running between the EW scale and the $B$ meson scale is performed, our results can be used systematically to test the sensitivity of $B$ physics observables on the dimension six operators.

\acknowledgments
%
We thank M.\ Gonz\'{a}lez-Alonso, U.\ Haisch and D.\ Straub for useful correspondence.
We also thank K.\ Teppei for confirming missing factors of 2 in eqs.~(\ref{eqn:1loopBBarC1}-\ref{eqn:1loopBBarC5}) in earlier versions and the journal version.
J.A., M.F.\ and C.G.\ acknowledge the support from the Swiss National Science Foundation.
A.C.~is supported by a Marie Curie Intra-European Fellowship of the European Community's 7th Framework Programme under contract number PIEF-GA-2012-326948.

\appendix

\section{Dimension-six operators in the mass basis}
\label{sec:appop}
Here we explicitly relate the Wilson coefficients of the gauge invariant operators in the interaction basis to the mass basis. This translation is necessary, if the results obtained in this article have to be related to a UV complete model, where the interaction basis is specified.
For the notation, we refer the reader to the original paper in ref.~\cite{GIMR}.

\small
\begin{longtable}{c|l}
 \caption{Operators with quarks, gauge and/or Higgs bosons}\\
 \hline \hline
 Operator & Definition in the mass basis\\
 \endfirsthead
 \hline \hline
 Operator & Definition in the mass basis\\
 \endhead
 \hline\\[-2.5ex]
 $Q_{dB}$ &
 $ \tilde{C}_{dB}^{ij}
  \left[ V_{ki} \;
    \bar{u}_L^k \sigma^{\mu\nu} d_R^j \varphi^+
    +\bar{d}_L^i \sigma^{\mu\nu} d_R^j\;
    \left( \frac{v+h+i\varphi^0}{\sqrt{2}} \right)
  \right]  B_{\mu\nu}$\\
  &$\tilde{C}_{dB}^{ij} =
   C_{dB}^{mn} S^{d\dagger}_{L\, im} S^d_{R\, nj}$
   \\[0.5ex] \hline \\[-2.5ex]
  $Q_{dW}$ &
  $\tilde{C}_{dW}^{ij}
  \left[ V_{ki} \;
    \bar{u}_L^k \sigma^{\mu\nu} d_R^j \varphi^+
    -\bar{d}_L^i \sigma^{\mu\nu} d_R^j\;
    \left( \frac{v+h+i\varphi^0}{\sqrt{2}} \right)
  \right]   W_{\mu\nu}^3 + \dots$\\
  &$\tilde{C}_{dW}^{ij} =
   C_{dW}^{mn} S^{d\dagger}_{L\, im} S^d_{R\, nj}$
   \\[0.5ex] \hline \\[-2.5ex]
 $Q_{uB}$ &
 $ \tilde{C}_{uB}^{ij}
  \left[     \bar{u}_L^i \sigma^{\mu\nu} u_R^j\;
    \left( \frac{v+h-i\varphi^0}{\sqrt{2}} \right)
    -V_{ik}^* \; \bar{d}_L^k \sigma^{\mu\nu} u_R^j \varphi^-
  \right]  B_{\mu\nu}$\\
  &$\tilde{C}_{uB}^{ij} =
   C_{uB}^{mn} S^{u\dagger}_{L\, im} S^u_{R\, nj}$
   \\[0.5ex] \hline \\[-2.5ex]
  $Q_{uW}$ &
  $\tilde{C}_{uW}^{ij}
  \left[     \bar{u}_L^i \sigma^{\mu\nu} u_R^j\;
    \left( \frac{v+h-i\varphi^0}{\sqrt{2}} \right)
    +V_{ik}^* \; \bar{d}_L^k \sigma^{\mu\nu} u_R^j \varphi^-
  \right]   W_{\mu\nu}^3 + \dots$\\
  &$\tilde{C}_{uW}^{ij} =
   C_{uW}^{mn} S^{u\dagger}_{L\, im} S^u_{R\, nj}$
   \\[0.5ex] \hline \\[-2.5ex]
  $Q_{dG}$ &
  $\tilde{C}_{dG}^{ij}
  \left[ V_{ki} \;  \bar{u}_L^k \sigma^{\mu\nu} T^A d_R^j \varphi^+
    +\bar{d}_L^i \sigma^{\mu\nu} T^A d_R^j\;
    \left( \frac{v+h+i\varphi^0}{\sqrt{2}} \right)
  \right]  G_{\mu\nu}^A$\\
  &$\tilde{C}_{dG}^{ij} =
   C_{dG}^{mn} S^{d\dagger}_{L\, im} S^d_{R\, nj}$
   \\[0.5ex] \hline \\[-2.5ex]
  $Q_{uG}$ &
  $\tilde{C}_{uG}^{ij}
  \left[
    \bar{u}_L^i \sigma^{\mu\nu} T^A u_R^j\;
    \left( \frac{v+h-i\varphi^0}{\sqrt{2}} \right)
    -V_{ik}^* \;  \bar{d}_L^k \sigma^{\mu\nu} T^A u_R^j \, \varphi^-
  \right]  G_{\mu\nu}^A$\\
  &$\tilde{C}_{uG}^{ij} =
   C_{uG}^{mn} S^{u\dagger}_{L\, im} S^u_{R\, nj}$
   \\[0.5ex] \hline \\[-2.5ex]
  $Q_{\varphi q}^{(1)}$ &
  $\tilde{C}_{\varphi q}^{(1) \, ij}
  \left( \varphi^\dagger i \overset{\leftrightarrow}{D}_\mu \varphi \right)
  \left( V_{mi} V^*_{nj}  \, \bar{u}_L^m \gamma^\mu u_L^n +\bar{d}_L^i \gamma^\mu d_L^j \right)$\\
  &$\tilde{C}_{\varphi q}^{(1) \, ij} =
   C_{\varphi q}^{(1) \, mn}    S^{d\dagger}_{L \, im} S^d_{L \,nj}$
   \\[0.5ex] \hline \\[-2.5ex]
  $Q_{\varphi q}^{(3)}$ &
  $\tilde{C}_{\varphi q}^{(3) \, ij}
  \left( \varphi^\dagger i \overset{\leftrightarrow}{D}_\mu^1 \varphi \right)
  \left(  V_{mi} \, \bar{u}_L^m \gamma^\mu d_L^j +V^*_{nj} \, \bar{d}_L^i \gamma^\mu u_L^n\right)
  + \dots $\\
  &$ \tilde{C}_{\varphi q}^{(3) \, ij} =
   C_{\varphi q}^{(3) \, mn}    S^{d\dagger}_{L \, im} S^d_{L \,nj}$
   \\[0.5ex] \hline \\[-2.5ex]
  $Q_{\varphi d}$ &
  $\tilde{C}_{\varphi d}^{ij}
  \left( \varphi^\dagger i \overset{\leftrightarrow}{D}_\mu \varphi \right)
  \left( \bar{d}_R^i \gamma^\mu d_R^j \right)$\\
  &$  \tilde{C}_{\varphi d}^{ij} =
  C_{\varphi d}^{mn}    S^{d\dagger}_{R \, im} S^d_{R \,nj}$
   \\[0.5ex] \hline \\[-2.5ex]
  $Q_{\varphi u}$ &
  $\tilde{C}_{\varphi u}^{ij}
  \left( \varphi^\dagger i \overset{\leftrightarrow}{D}_\mu \varphi \right)
  \left( \bar{u}_R^i \gamma^\mu u_R^j \right)$\\
  &$ \tilde{C}_{\varphi u}^{ij} =
   C_{\varphi u}^{mn} S^{u\dagger}_{R \, im} S^u_{R \,nj}$
   \\[0.5ex] \hline \\[-2.5ex]
  $Q_{\varphi ud}$ &
   $\tilde{C}_{\varphi ud}^{ij}
  \, \left( \tilde{\varphi}^\dagger  i D_\mu \varphi \right)
  \left( \bar{u}_R^i \gamma^\mu d_R^j \right)$\\
  &$ \tilde{C}_{\varphi ud}^{ij} =
   C_{\varphi ud}^{mn} S^{u\dagger}_{R \, im} S^d_{R \,nj}$
   \\[0.5ex] \hline \\[-2.5ex]
   $Q_{d\varphi}$ &
   $\tilde{C}_{d\varphi}^{ij} \left( \varphi^\dagger \varphi \right)
   \left[ V_{mi} \bar{u}_L^m d_R^j \varphi^+
   + \bar{d}_L^i d_R^j \left( \frac{v+h+i\varphi^0}{\sqrt{2}} \right)\right]$ \\
   &$ \tilde{C}_{d \varphi }^{ij} =
   C_{d \varphi }^{mn} S^{d\dagger}_{L \, im} S^d_{R \,nj}$
   \\[0.5ex]  \hline
\end{longtable}

\begin{longtable}{c|l}
 \caption{Four-fermion operators with four quarks}\\
 \hline \hline
 Operator & Definition in the mass basis\\
 \endfirsthead
 \hline \hline
 Operator & Definition in the mass basis\\
 \endhead
 \hline\\[-2.5ex]
 $Q_{qq}^{(1)}$ &
 $\tilde{C}_{qq}^{(1) \, ijkl}
  \left( V_{mi} V^*_{nj}\, \bar{u}_L^m \gamma^\mu u_L^n + \bar{d}_L^i \gamma^\mu d_L^j \right)
  \left( V_{mk} V^*_{nl}\, \bar{u}_L^m \gamma_\mu u_L^n + \bar{d}_L^k \gamma_\mu d_L^l \right)$ \\
  & $\tilde{C}_{qq}^{(1) \, ijkl} =  C_{qq}^{(1) \, pqrs} S^{d\dagger}_{L\, ip} S^{d}_{L\, qj}
  S^{d\dagger}_{L\, kr}S^{d}_{L\, sl}$\\[0.5ex] \hline \\[-2.5ex]
  $Q_{qq}^{(3)}$ &
  $\tilde{C}_{qq}^{(3) \, ijkl}
  \left( V_{mi} V^*_{nj}\,\bar{u}_L^m \gamma^\mu u_L^n - \bar{d}_L^i \gamma^\mu d_L^j \right)
  \left( V_{mk} V^*_{nl} \,\bar{u}_L^m \gamma_\mu u_L^n - \bar{d}_L^k \gamma_\mu d_L^l \right)
  +\dots$ \\
  &$\tilde{C}_{qq}^{(3) \, ijkl} =
  C_{qq}^{(3) \, pqrs} S^{d\dagger}_{L\, ip} S^{d}_{L\, qj}
  S^{d\dagger}_{L\, kr}S^{d}_{L\, sl}$
  \\[0.5ex] \hline \\[-2.5ex]
  $ Q_{qd}^{(1)}$ &
  $\tilde{C}_{qd}^{(1) \, ijkl}
  \left( V_{mi} V^*_{nj}\, \bar{u}_L^m \gamma^\mu u_L^n + \bar{d}_L^i \gamma^\mu d_L^j \right)
  \left(\bar{d}_R^k \gamma_\mu d_R^l \right)$ \\
 & $\tilde{C}_{qd}^{(1) \, ijkl} =
 C_{qd}^{(1) \, pqrs} S^{d\dagger}_{L\, ip} S^{d}_{L\, qj}
 S^{d\dagger}_{R \,kr}S^{d}_{R \,sl}$
 \\[0.5ex] \hline  \\[-2.5ex]
  $Q_{qd}^{(8)}$ &
  $\tilde{C}_{qd}^{(8) \, ijkl}
  \left( V_{mi} V^*_{nj}\,\bar{u}_L^m\gamma^\mu\, T^A u_L^n +\bar{d}_L^i \gamma^\mu\, T^Ad_L^j \right)
  \left(\bar{d}_R^k \gamma_\mu T^A d_R^l \right)$ \\
  &$\tilde{C}_{qd}^{(8) \, ijkl} =
  C_{qd}^{(8) \, pqrs} S^{d\dagger}_{L\, ip} S^{d}_{L\, qj}
  S^{d\dagger}_{R \,kr}S^{d}_{R \,sl}$
  \\[0.5ex] \hline  \\[-2.5ex]
  $ Q_{qu}^{(1)}$ &
  $\tilde{C}_{qu}^{(1) \, ijkl}
  \left( V_{mi} V^*_{nj}\, \bar{u}_L^m \gamma^\mu u_L^n + \bar{d}_L^i \gamma^\mu d_L^j \right)
  \left(\bar{u}_R^k \gamma_\mu u_R^l \right)$ \\
 & $\tilde{C}_{qu}^{(1) \, ijkl} =
 C_{qu}^{(1) \, pqrs} S^{d\dagger}_{L\, ip} S^{d}_{L\, qj}
 S^{u\dagger}_{R \,kr}S^{u}_{R \,sl}$
 \\[0.5ex] \hline \\[-2.5ex]
  $Q_{qu}^{(8)}$ &
  $\tilde{C}_{qu}^{(8) \, ijkl}
  \left( V_{mi} V^*_{nj}\,\bar{u}_L^m\gamma^\mu\, T^A u_L^n +\bar{d}_L^i \gamma^\mu\, T^Ad_L^j \right)
  \left(\bar{u}_R^k \gamma_\mu T^A u_R^l \right)$ \\
  &$\tilde{C}_{qu}^{(8) \, ijkl} =
  C_{qu}^{(8) \, pqrs} S^{d\dagger}_{L\, ip} S^{d}_{L\, qj}
  S^{u\dagger}_{R \,kr}S^{u}_{R \,sl}$
  \\[0.5ex] \hline \\[-2.5ex]
  $Q_{ud}^{(1)}$ &
  $ \tilde{C}_{ud}^{(1)\, ijkl}  \left( \bar{u}_R^i \gamma^\mu u_R^j \right)
  \left(\bar{d}_R^k \gamma_\mu d_R^l \right)$ \\
 &  $\tilde{C}_{ud}^{(1)\, ijkl} =
  C_{ud}^{(1)\, pqrs} S^{u\dagger}_{R\, ip} S^{u}_{R\, qj}
  S^{d\dagger}_{R \,kr}S^{d}_{R \,sl}$
  \\ [0.5ex] \hline \\[-2.5ex]
  $Q_{ud}^{(8)}$ &
  $ \tilde{C}_{ud}^{(8)\, ijkl}  \left( \bar{u}_R^i T^A \gamma^\mu u_R^j \right)
  \left(\bar{d}_R^k T^A \gamma_\mu d_R^l \right)$ \\
 &  $\tilde{C}_{ud}^{(8)\, ijkl} =
  C_{ud}^{(8)\, pqrs} S^{u\dagger}_{R\, ip} S^{u}_{R\, qj}
  S^{d\dagger}_{R \,kr}S^{d}_{R \,sl}$
  \\ [0.5ex] \hline \\[-2.5ex]
  $Q_{dd}$ &
  $ \tilde{C}_{dd}^{ijkl}  \left( \bar{d}_R^i \gamma^\mu d_R^j \right)
  \left(\bar{d}_R^k \gamma_\mu d_R^l \right)$ \\
 &  $\tilde{C}_{dd}^{ijkl} =
  C_{dd}^{pqrs} S^{d\dagger}_{R\, ip} S^{d}_{R\, qj}
  S^{d\dagger}_{R \,kr}S^{d}_{R \,sl}$
  \\ [0.5ex] \hline \\[-2.5ex]
  $Q_{uu}$ &
  $ \tilde{C}_{uu}^{ijkl}  \left( \bar{u}_R^i \gamma^\mu u_R^j \right)
  \left(\bar{u}_R^k \gamma_\mu u_R^l \right)$ \\
 &  $\tilde{C}_{uu}^{ijkl} =
  C_{uu}^{pqrs} S^{u\dagger}_{R\, ip} S^{u}_{R\, qj}
  S^{u\dagger}_{R \,kr}S^{u}_{R \,sl}$
  \\ [0.5ex] \hline \\[-2.5ex]
  $Q_{quqd}^{(1)}$ &
  $\tilde{C}^{(1)\, ijkl}_{quqd} \left[
  \left( \bar{u}_L^i u_R^j \right)\left( \bar{d}_L^k d_R^l \right)
  -V^*_{im} V_{nk}\left( \bar{d}_L^m u_R^j \right)\left( \bar{u}_L^n d_R^l \right)
  \right]$ \\
  & $\tilde{C}_{quqd}^{(1) \, ijkl} =
  C_{quqd}^{(1) \, pqrs} S^{u\dagger}_{L\, ip} S^{u}_{R\, qj}
  S^{d\dagger}_{L \,kr}S^{d}_{R \,sl}$
  \\[0.5ex] \hline \\[-2.5ex]
  $Q_{quqd}^{(8)}$ &
  $\tilde{C}^{(8)\, ijkl}_{quqd} \left[
  \left( \bar{u}_L^i T^A u_R^j \right)\left( \bar{d}_L^k T^A d_R^l \right)
  -V^*_{im} V_{nk}\left( \bar{d}_L^m T^A u_R^j \right)\left( \bar{u}_L^n T^A d_R^l \right)
  \right]$ \\
  & $\tilde{C}_{quqd}^{(8) \, ijkl} =
  C_{quqd}^{(8) \, pqrs} S^{u\dagger}_{L\, ip} S^{u}_{R\, qj}
  S^{d\dagger}_{L \,kr}S^{d}_{R \,sl}$
  \\[0.5ex] \hline 
\end{longtable}
\begin{longtable}{c|l}
 \caption{Four-fermion operators with two quarks and two leptons}\\
 \hline \hline
 Operator & Definition in the mass basis\\
 \endfirsthead
 \hline \hline
 Operator & Definition in the mass basis\\
 \endhead
 \hline\\[-2.5ex]
 $Q^{(1)}_{\ell q}$ &
 $\tilde{C}^{(1) \, ijkl}_{\ell q}
 \left( \bar{\nu}\, _L^i \gamma^\mu \nu\, _L^j +
  \bar{e}\, _L^i \gamma^\mu e\, _L^j  \right)
  \left( V_{m k}\, V^*_{n l} \, \bar{u}\, _L^{m} \gamma_\mu u\, _L^{n}  +   \, \bar{d}\, _L ^{k} \gamma_\mu d \, _L^{l} \right)$ \\
 &$\tilde{C}^{(1) \, ijkl}_{\ell q} =  C^{(1) \, ijmn}_{\ell q} S^{d\dagger}_{L \, km} S^d_{L \,nl}$
 \\ [0.5ex] \hline \\[-2.5ex]
 $Q^{(3)}_{\ell q}$ &
 $\tilde{C}^{(3) \, ijkl}_{\ell q}
 \left( \bar{\nu}\, _L^i \gamma^\mu \nu\, _L^j
  -\bar{e}\, _L^i \gamma^\mu e\, _L^j  \right)
  \left( V_{mk}\, V^*_{nl} \,\bar{u}\, _L^{m} \gamma_\mu u\, _L^{n}
  - \bar{d}\, _L ^{k} \gamma_\mu d \, _L^{l} \right) \cdots $\\
 & $\tilde{C}^{(3) \, ijkl}_{\ell q} =
  C^{(3) \, ijmn}_{\ell q} S^{d\dagger}_{L \, km} S^d_{L \, nl}$
  \\ [0.5ex] \hline  \\[-2.5ex]
 $Q_{eu}$ &
 $\tilde{C}^{ijkl}_{eu}
  \left( \bar{e}\, _R^i \gamma^\mu e\, _R^j \right)
  \left( \bar{u}\, _R^k \gamma_\mu u\, _R^l \right)$\\
 &$\tilde{C}^{ijkl}_{eu} =
  C^{ijmn}_{eu} S^{u\dagger}_{R \, km } S^u_{R \, n l}$\\[0.5ex] \hline \\[-2.5ex]
 $Q_{ed}$&
 $\tilde{C}_{ed}^{ijkl}
  \left( \bar{e}\, _R^i \gamma^\mu e\, _R^j \right)
  \left( \bar{d}\, _R^k \gamma_\mu d\, _R^l \right)$\\
&$\tilde{C}^{ijkl}_{ed} =
  C^{ijmn}_{ed} S^{d\dagger}_{R \, km} S^d_{R \, nl}$\\[0.5ex] \hline \\[-2.5ex]
 $Q_{\ell u}$&
 $\tilde{C}^{ijkl}_{\ell u}
  \left( \bar{\nu}\, _L^i \gamma^\mu \nu\, _L^j+  \bar{e}\, _L^i \gamma^\mu e\, _L^j \right)
  \left( \bar{u}\, _R^k \gamma_\mu u\, _R^l \right)$\\
& $\tilde{C}^{ijkl}_{\ell u} =  C^{ijmn}_{\ell u}S^{u\dagger}_{R \, km} S^u_{R \, nl}$\\[0.5ex] \hline \\[-2.5ex]
%
 $Q_{\ell d}$&
 $\tilde{C}^{ijkl}_{\ell d}   \left( \bar{\nu}\, _L^i \gamma^\mu \nu\, _L^j+   \bar{e}\, _L^i \gamma^\mu e\, _L^j \right)
 \left( \bar{d}\, _R^k \gamma_\mu d\, _R^l \right)$\\
&$\tilde{C}^{ijkl}_{\ell d} =  C^{ijmn}_{\ell d}S^{d\dagger}_{R \, km} S^d_{R \, nl}$\\[0.5ex] \hline \\[-2.5ex]
%
 $Q_{qe}$&
 $\tilde{C}^{ijkl}_{qe}  \left( V_{mi}\, V^*_{nj} \,\bar{u}\, _L^{m } \gamma_\mu u\, _L^{n }  + \bar{d} \,_L^{i} \gamma_\mu d \,_L^{j} \right)
 \left( \bar{e}\, _R^k \gamma^\mu e\, _R^l \right)$\\
 &$\tilde{C}^{ijkl}_{qe} =  C^{mnkl}_{qe}S^{d\dagger}_{L \, im} S^d_{L \, nj}$\\[0.5ex] \hline \\[-2.5ex]
 $Q_{\ell edq}$&
 $\tilde{C}^{ijkl}_{\ell edq}
  \left[  V^*_{nl}\left(\bar{\nu}\, _L^i e\, _R^j \right) \, \left( \bar{d}\, _R^k u_L^{n }\right)
  + \left(\bar{e}\, _L^i  e\, _R^j\right) \,\left( \bar{d} \,_R^k d_L^{l}\right)\right]$\\
&$\tilde{C}^{ijkl}_{\ell edq} =
C^{ijmn}_{\ell edq}S^{d\dagger}_{R \, km} S^d_{L \, nl}$\\[0.5ex] \hline \\[-2.5ex]
 $Q^{(1)}_{\ell equ}$&
 $\tilde{C}^{(1) \, ijkl}_{\ell equ}    \left[   V^*_{km}\left(\bar{\nu}\, _L^i e\, _R^j \right)     \,  \left( \bar{d} \,_L^{m } u_R^l\right)
 -\left(\bar{e}\, _L^i  e\, _R^j\right)   \,\left( \bar{u}_L^k u_R^{l}\right)\right]$\\
 & $\tilde{C}^{(1) \, ijkl}_{\ell equ} =
 C^{(1) \, ijmn}_{\ell equ}S^{u\dagger}_{L \, km} S^u_{R \, nl}$
 \\[0.5ex] \hline  \\[-2.5ex]
 $Q^{(3)}_{\ell equ}$&
 $\tilde{C}^{(3) \, ijkl}_{\ell equ}  \left[     V^*_{km}\left(\bar{\nu}\, _L^i \sigma^{\mu\nu}e\, _R^j \right)
    \,  \left( \bar{d} \,_L^{m } \sigma_{\mu\nu}u_R^l\right)
  -\left(\bar{e}\, _L^i \sigma^{\mu\nu} e\, _R^j\right)
\,\left( \bar{u}_L^k \sigma_{\mu\nu}u_R^{l}\right)\right]$\\
&$\tilde{C}^{(3) \, ijkl}_{\ell equ} =
  C^{(3) \, ijmn}_{\ell equ}S^{u\dagger}_{L \, km} S^u_{R \, nl}$\\[0.5ex] \hline 
\end{longtable}

\label{Bibliography}
\bibliographystyle{JHEP}
 \footnotesize
\bibliography{BIB}
\end{document}